\documentclass[]{aastex61}

\usepackage{amsmath}
\usepackage{ulem}

\DeclareMathOperator\erf{erf}

\begin{document}

\title{ALMA resolves \ion{C}{1} emission from the $\beta$~Pictoris debris disk}

\correspondingauthor{Gianni Cataldi}
\email{cataldi@naoj.org}

\author[0000-0002-2700-9676]{Gianni Cataldi}
\altaffiliation{International Research Fellow of Japan Society for the Promotion of Science (Postdoctoral Fellowships for Research in Japan (Standard)).}
\affil{Subaru Telescope, National Astronomical Observatory of Japan, 650 North A’ohoku Place, Hilo, HI 96720, USA}
\affil{Konkoly Observatory, Research Centre for Astronomy and Earth Sciences, Hungarian Academy of Sciences, Konkoly-Thege Mikl\'os \'ut 15--17, 1121 Budapest, Hungary}
\affil{AlbaNova University Centre, Stockholm University, Department of Astronomy, Stockholm, Sweden}
\affil{Stockholm University Astrobiology Centre, Stockholm, Sweden}

\author[0000-0002-7201-7536]{Alexis Brandeker}
\affil{AlbaNova University Centre, Stockholm University, Department of Astronomy, Stockholm, Sweden}
\affil{Stockholm University Astrobiology Centre, Stockholm, Sweden}

\author{Yanqin Wu}
\affil{Department of Astronomy \& Astrophysics, University of Toronto, Ontario, Canada}

\author{Christine Chen}
\affil{Space Telescope Science Institute, 3700 San Martin Dr., Baltimore, MD 21218, USA}
\affil{Department of Physics and Astronomy, The Johns Hopkins University, 3400 North Charles Street, Baltimore, MD, 21218, USA}

\author{William Dent}
\affil{ALMA/ESO, Alonso de Cordova 3107, Santiago, Chile}

\author{Bernard L. de Vries}
\affil{AlbaNova University Centre, Stockholm University, Department of Astronomy, Stockholm, Sweden}
\affil{Stockholm University Astrobiology Centre, Stockholm, Sweden}
\affil{Scientific Support Office, Directorate of Science, European Space Research and Technology Centre (ESA/ESTEC), Keplerlaan 1, 2201 AZ Noordwijk, The Netherlands}

\author{Inga Kamp}
\affil{Kapteyn Astronomical Institute, University of Groningen, The Netherlands}

\author[0000-0002-7135-2439]{Ren\'e Liseau}
\affil{Department of Space, Earth and Environment, Chalmers University of Technology, Onsala Space Observatory, 439 92 Onsala, Sweden}

\author[0000-0003-3747-7120]{G\"oran Olofsson}
\affil{AlbaNova University Centre, Stockholm University, Department of Astronomy, Stockholm, Sweden}
\affil{Stockholm University Astrobiology Centre, Stockholm, Sweden}

\author{Eric Pantin}
\affil{Laboratoire AIM, UMR 7158, CEA/DSM---CNRS---Université\'e Paris Diderot, IRFU/SAp, 91191 Gif sur Yvette, France}

\author[0000-0002-2989-3725]{Aki Roberge}
\affil{Exoplanets and Stellar Astrophysics Lab, NASA Goddard Space Flight Center, Greenbelt, MD 20771, USA}

% \author[0000-0002-0786-7307]{Greg J. Schwarz}
% \affil{American Astronomical Society \\
% 2000 Florida Ave., NW, Suite 300 \\
% Washington, DC 20009-1231, USA}

% \author{August Muench}
% \affiliation{American Astronomical Society \\
% 2000 Florida Ave., NW, Suite 300 \\
% Washington, DC 20009-1231, USA}
% \collaboration{(AAS Journals Data Scientists collaboration)}

% \author{Butler Burton}
% \affiliation{National Radio Astronomy Observatory}
% \affiliation{AAS Journals Associate Editor-in-Chief}
% \nocollaboration

% \author{Amy Hendrickson}
% \altaffiliation{Creator of AASTeX v6.1}
% \affiliation{TeXnology Inc.}
% \collaboration{(LaTeX collaboration)}

% AAS Journals, the Astrophysical Journal (ApJ), the
% Astrophysical Journal Letters (ApJL), and Astronomical Journal (AJ), all
% have a 250 word limit for the abstract.  If you exceed this length the
% Editorial office will ask you to shorten it.
\begin{abstract}
The debris disk around $\beta$~Pictoris is known to contain gas. Previous ALMA observations revealed a CO belt at $\sim$85\,au with a distinct clump, interpreted as a location of enhanced gas production. Photodissociation converts CO into C and O within $\sim$50 years. We resolve \ion{C}{1} emission at 492\,GHz using ALMA and study its spatial distribution. \ion{C}{1} shows the same clump as seen for CO. This is surprising, as C is expected to quickly spread in azimuth. We derive a low C mass (between $5\times10^{-4}$ and $3.1\times10^{-3}$\,M$_\earth$), indicating that gas production started only recently (within $\sim$5\,000 years). No evidence is seen for an atomic accretion disk inwards of the CO belt, perhaps because the gas did not yet have time to spread radially. The fact that C and CO share the same asymmetry argues against a previously proposed scenario where the clump is due to an outward migrating planet trapping planetesimals in an resonance; nor can the observations be explained by an eccentric planetesimal belt secularly forced by a planet. Instead, we suggest that the dust and gas disks should be eccentric. Such a configuration, we further speculate, might be produced by a recent tidal disruption event. Assuming that the disrupted body has had a CO mass fraction of 10\%, its total mass would be $\gtrsim$3\,$M_\mathrm{Moon}$.
\end{abstract}


\section{Introduction}
In debris disk systems, the continuous collisional destruction of larger bodies such as comets or asteroids produces abundant amounts of dust, with the smallest grains quickly removed by radiation pressure \citep{Backman_Paresce_1993,Wyatt_2008}. A debris disk provides evidence that planetesimal-sized bodies were formed during the earlier protoplanetary phase \citep{Artymowicz_1997,Matthews_etal_2014} and gives us the opportunity to study the properties of the building blocks of planets. Studying these systems is therefore intimately linked to our efforts of understanding how planets form.

The debris disk around the young \citep[$23\pm3$\,Ma,][]{Mamajek_Bell_2014} A6V star \citep{Gray_etal_2006} $\beta$~Pictoris has been used as a laboratory to study the early evolution of planetary systems ever since its discovery by the Infrared Astronomical Satellite (IRAS) \citep{Aumann_1985}. \citet{Smith_Terrile_1984} obtained the first resolved image showing an edge-on disk. Since then, the properties of the dust disk have been extensively studied with observations at multiple wavelengths. Today, we know that the belt of parent bodies is located at $\sim$100\,au \citep{Dent_etal_2014} and that $\beta$~Pic hosts a giant planet \citep[e.g.][and references therein]{Chilcote_etal_2017} with a semi-major axis of $\sim$10\,au \citep{LecavelierdesEtangs_Vidal-Madjar_2016,Wang_etal_2016}, first imaged by \citet{Lagrange_etal_2009,Lagrange_etal_2010}.

Even before the dust disk was discovered, evidence from optical and ultraviolet (UV) absorption lines suggested the presence of gas around $\beta$~Pic \citep{Slettebak_1975,Slettebak_Carpenter_1983}. The $\beta$~Pic disk is thus part of a small sub-sample of debris disks where gas has been detected. Currently, there are about 20 such gaseous debris disks known \citep[e.g.][]{Redfield_2007,Moor_etal_2011,Roberge_etal_2014,Lieman-Sifry_etal_2016,Matra_etal_2017_Fomalhaut}.

The spatial distribution of the gas around $\beta$~Pic has been studied with resolved observations in the optical and recently with the Atacama Large Millimeter/submillimeter Array (ALMA). These data showed that the gas disk is radially extended (with some species traced out to several hundred au) and in Keplerian rotation \citep{Olofsson_etal_2001,Brandeker_etal_2004,Nilsson_etal_2012,Dent_etal_2014}. Besides this stable component, time-variable absorption features shifted with respect to the systemic velocity are attributed to exocomets evaporating in vicinity of the star \citep[e.g.][and references therein]{Vidal-Madjar_etal_1994,Kiefer_etal_2014}. This latter phenomenon has also been seen around a number of other (mostly young) A-type stars \citep[e.g.][and references therein]{Welsh_Montgomery_2015}.

Similarly to the dust, the gas in the $\beta$~Pic disk is thought to be continuously produced from the destruction of solid material rather than leftover from the protoplanetary phase. Evidence for such a secondary scenario comes for example from theoretical arguments on the dynamical lifetime of the gas \citep{Fernandez_etal_2006}. Also, models of the excitation of the CO 3--2 and 2--1 transitions observed by ALMA imply that not enough H$_2$ is present in the disk to shield CO from photodissociation over the lifetime of the disk, thus the necessity of a gas replenishment mechanism \citep{Matra_etal_2017}. Studying this secondary gas opens up the interesting possibility to constrain the composition of the parent bodies \citep[e.g.][]{Kral_etal_2016,Matra_etal_2017,Matra_etal_2018}.

To date, multiple metallic species such as C, O, Na, Al or Ca have been detected \citep[e.g.][]{Brandeker_etal_2004,Roberge_etal_2006,Brandeker_etal_2016}. Recently, the first detection of hydrogen was reported by \citet{Wilson_etal_2017}. CO remains the only molecule detected so-far \citep[e.g.][]{Dent_etal_2014,Matra_etal_2018}. The observed elemental abundances are strikingly different from solar abundances. While the abundance of H is much lower than solar \citep{Wilson_etal_2017}, C is highly overabundant with respect to other metals \citep{Roberge_etal_2006,Cataldi_etal_2014}. \citet{Fernandez_etal_2006} showed that the carbon overabundance provides a braking mechanism, preventing other metals that are strongly affected by radiation pressure (such as Na) from being quickly ejected from the system. Carbon also plays a crucial role in determining the excitation conditions of atomic fine-structure or molecular rotational transitions \citep[e.g.][]{Zagorovsky_etal_2010}. This is because in a secondary, hydrogen-depleted disk, collisional excitation is likely dominated by electrons, and ionisation of carbon is the main electron source. 

Using spectrally resolved observations of \ion{C}{2} with \textit{Herschel}\footnote{{\it Herschel} is an ESA space observatory with science instruments provided by European-led Principal Investigator consortia and with important participation from NASA.}/HIFI, the spatial distribution of carbon was constrained by \citet{Cataldi_etal_2014}. They found that most of the carbon gas is located at $\sim$100\,au, with tentatively more emission from the south-west (SW) side of the disk. At about the same time, ALMA spatially resolved CO~3--2 emission, revealing a clump of emission on the SW side of the disk at $\sim$85\,au \citep{Dent_etal_2014}. The CO clump coincides with a radial peak of the millimetre continuum \citep{Dent_etal_2014} and a clump seen at mid-IR wavelengths \citep{Telesco_etal_2005,Li_etal_2012}. Dissociation by interstellar UV photons limits the lifetime of CO in the disk to significantly less than an orbit \citep{Visser_etal_2009}. It is thus clear that CO needs to be produced continuously, and is a source for both C and O. This might provide a natural explanation for the observed super-solar abundance of C with respect to metals such as Na \citep{Xie_etal_2013}. The CO itself is believed to be derived from the destruction of volatile-rich, cometary bodies, where the clump corresponds to a location of increased collision rate and thus gas production. Several hypothesis have been put forward to explain the existence of the clump. Firstly, it could be the location of a giant collision. The clump then results from the fact that the orbits of the collision debris always all go through the collision point \citep{Jackson_etal_2014}. Secondly, the clump could be due to cometary bodies trapped in a resonance with an outward migrating, yet unseen giant planet \citep{Wyatt_2003,Wyatt_2006}. Using ALMA follow-up observations of the CO~2--1 transition at higher resolution, \citet{Matra_etal_2017} dismissed the giant collision scenario based on the large radial extent of the clump. Thirdly, \citet{Nesvold_Kuchner_2015} proposed that the interaction between a spiral density wave and a vertical displacement wave, both induced by $\beta$~Pic~b, can produce an azimuthaly asymmetric collision rate.

Assuming that indeed all C and O is derived from the dissociation of CO, \citet{Kral_etal_2016} modelled the hydrodynamical evolution of carbon and oxygen in the disk. They assumed that the produced atomic gas is subject to the magneto-rotational instability (MRI) and predict an atomic accretion and decretion disk \citep{Xie_etal_2013,Kral_Latter_2016}.

\ion{C}{1} was previously seen in absorption against $\beta$~Pic \citep{Vidal-Madjar_etal_1994,Jolly_etal_1998,Roberge_etal_2000}. Recently, \citet{Higuchi_etal_2017} presented the first observation of \ion{C}{1} in emission. In this paper, we present the first spatially resolved observations of \ion{C}{1}, revealing its distribution in the disk. Our paper is organised as follows. We describe the observations in section \ref{sec:observations} and present the results in section \ref{sec:results}. In section \ref{sec:modelling}, we present simple gas emission models to study the total mass and spatial distribution of the carbon gas. We discuss our results in section \ref{sec:discussion} and conclude in section \ref{sec:conclusion}.

\section{Observations and data reduction}\label{sec:observations}
%the integration times are from au.timeOnSource, where I use the latency-removed times
%for baselines, I used au.getBaselineStats and au.getBaselineLengths
%for pwv, I used au.getMedianPWV
%for baseline corresponding to the taper, I used au.angularScaleBaseline
We observed the $\beta$~Pic disk using band 8 receivers of ALMA on December 19, 2015 during the ALMA cycle 3 Early Science campaign (project ID 2013.1.00459.S, PI: Brandeker). The observations were split into two execution blocks (EBs). The total integration time was 2.1\,h (with 1.2\,h on $\beta$~Pic) and the median precipitable water vapor (pwv) was 0.6\,mm with standard deviation of 0.1\,mm. The array consisted of 36 antennas arranged in a hybrid configuration containing both short and long baselines ranging from 15\,m to 6.3\,km. In principle, we are thus sensitive to angular scales between $\sim$0.02\arcsec\ and $\sim$5\arcsec\ on the sky\footnote{see ALMA Cycle 6 Technical Handbook, section 3.6 (Spatial Filtering), \url{https://almascience.nao.ac.jp/documents-and-tools/cycle6/alma-technical-handbook}}. However, the visibilities from the longer baselines were affected by large atmospheric phase fluctuations and therefore flagged during the calibration process (see below). The observations were executed as a mosaic to obtain uniform sensitivity over the whole disk. One pointing was centred on the star and two additional pointings were centred at $\pm6\arcsec$ along the position angle of the disk. The size of the primary beam is 11.8\arcsec. Antenna elevations varied between 27\arcdeg\ and 60\arcdeg.

We placed three spectral windows, each with 1920 channels and a channel spacing of 488\,kHz (total bandwidth 937.5\,MHz), onto the following transitions: \ion{C}{1}~$^3$P$_1$--$^3$P$_0$ at 492.16\,GHz, CS~10--9 at 489.75\,GHz and SiO~11--10 at 477.50\,GHz. For \ion{C}{1}, this corresponds to a channel spacing of 0.30\,km\,s$^{-1}$ and an effective spectral resolution\footnote{see ALMA Cycle 6 Technical Handbook, Table 5.2} of 0.34\,km\,s$^{-1}$ (spectral averaging factor $N=2$). In addition, a fourth spectral window with 128 channels and a total bandwidth of 1875\,MHz was placed at 479\,GHz in order to observe the dust continuum.

The data were calibrated using \textsc{CASA} 4.7.0 \citep{McMullin_etal_2007}. We performed standard water vapour radiometre (WVR) calibration and system temperature corrections. The following calibration sources were observed: J0522-3627 (bandpass), J0538-4405 (phase), J0519-4546 and J0538-4405 (flux). However, we discarded the data from the two flux calibrators and instead used J0522-3627 to calibrate both bandpass and flux because this latter source had a significantly better measurement of its absolute flux in the ALMA Calibrator Source Catalogue.

The antenna time-dependent gain calibration solutions were adversely affected by the high atmospheric phase fluctuations on the longer baselines. We therefore flagged all baselines longer than 2\,km, effectively removing one third of the baselines. In addition, two bad antennas (DA44 and DA54) were also flagged. This allowed us to derive acceptable calibration solutions.

For the spectral windows placed onto emission lines, we performed continuum subtraction using the \texttt{uvcontsub} task within \textsc{CASA}. We then imaged the visibilities using the \texttt{CLEAN} task within \textsc{CASA}. In order to increase our surface brightness sensitivity, we applied a taper of 1\arcsec, thus significantly reducing the contribution of the remaining long baselines (at 492\,GHz, an angular scale of 1\arcsec corresponds to a baseline length of 126\,m). We also produced a continuum image using \texttt{CLEAN} with the same taper, combining all spectral windows except the one centred onto CS~10--9, which is in a region of bad atmospheric transmission and therefore particularly noisy.
%Among the unflagged baselines (those shorter than 2\,km), about one third are shorter than 126\,m, i.e., about two thirds are significantly attenuated by the taper.

Because of the low antenna elevations and the suboptimal array configuration, the sensitivity of the data is significantly worse than requested. Consequently, the data initially did not pass Quality Assurance 2 (QA2). However, we still decided to publish the data as the signal-to-noise ratio (SNR) is sufficient to provide new information on the $\beta$~Pic system.

\section{Results}\label{sec:results}
\subsection{Line emission}\label{sec:results_line_emission}
\begin{figure}
\plottwo{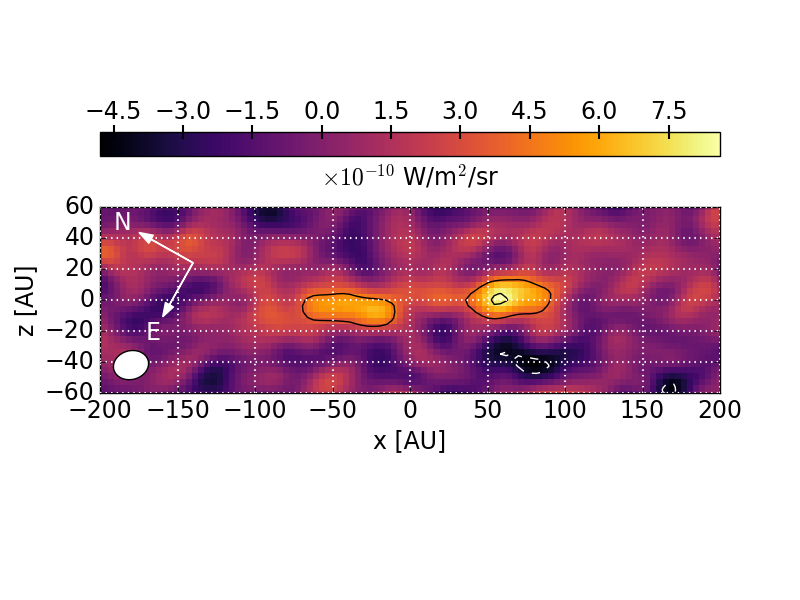}{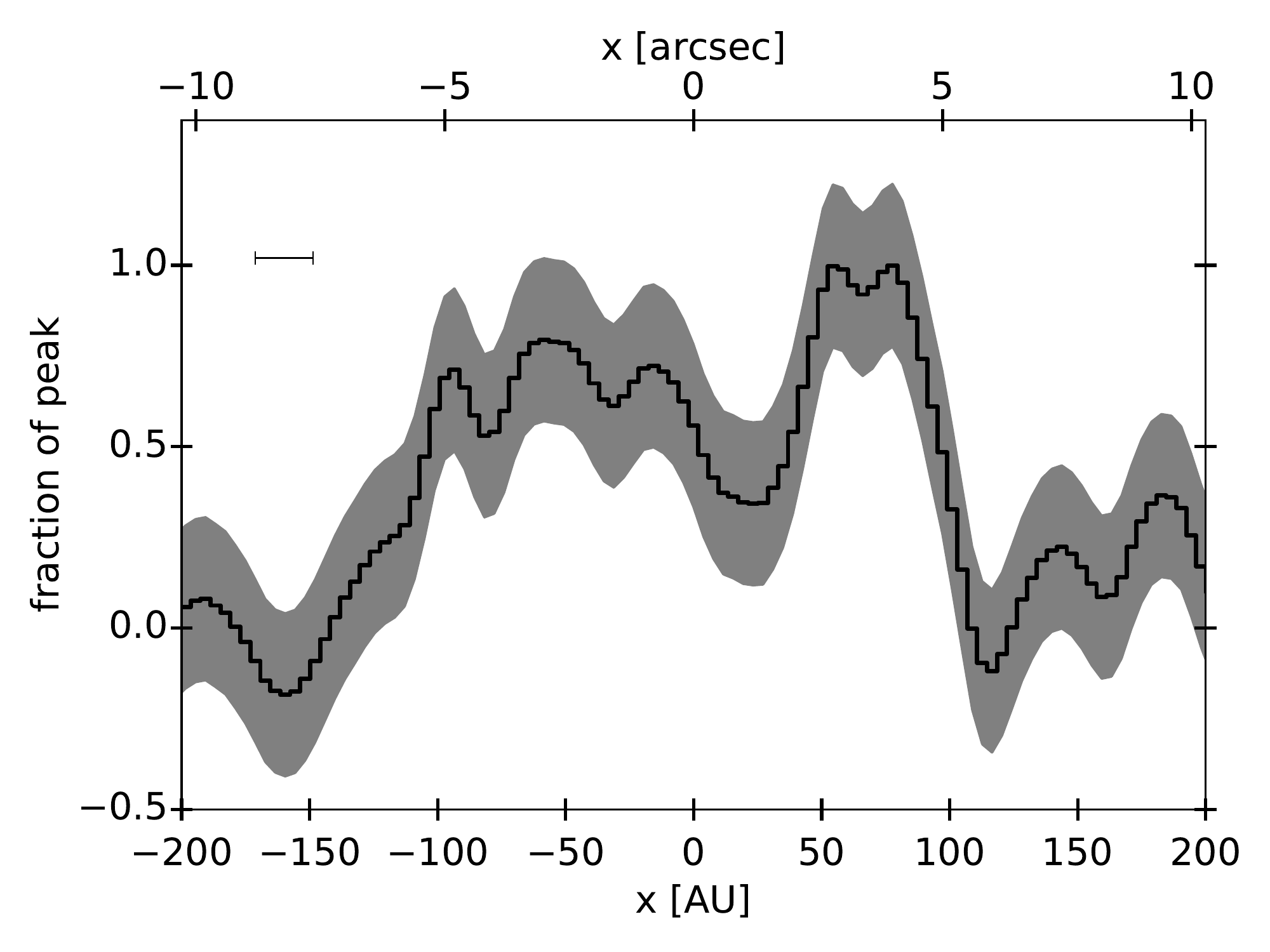}
\caption{\textit{Left}: Moment 0 of the \ion{C}{1} 492\,GHz emission from the $\beta$~Pic disk. Contours are drawn at intervals of 3$\sigma$, with negative contours drawn as dotted lines. The beam is illustrated as white ellipse in the lower left. The dashed rectangle illustrates the region used to measure the total flux. \textit{Right}: Emission profile along the x-axis of the moment 0 map, normalised to the peak value. The gray shaded area illustrates the $\pm 1\sigma$ error interval. The symbol in the upper left shows the projection of the beam onto the x-axis.\label{fig:mom0_xprofile}}
\end{figure}

%for the flux measurements and emission profiles, I used the methods working directly on the data cube (not on the mom0)
The CS~10--9 and SiO~11-10 lines remained undetected. We detect and resolve \ion{C}{1}~$^3$P$_1$--$^3$P$_0$ emission. Figure \ref{fig:mom0_xprofile} (left) shows the moment 0 map, produced by integrating the data cube along the spectral axis within $\pm6$\,km\,s$^{-1}$ \citep[with respect to the systemic velocity of the star, assumed to be $v_\mathrm{heliocentric}=20.5$\,km\,s$^{-1}$,][]{Brandeker_2011}. The figure has been rotated to align the horizontal direction with the main dust disk \citep[position angle of $+29.3\arcdeg$,][]{Lagrange_etal_2012}. The beam size is 1.18\arcsec$\times$0.95\arcsec (23\,au$\times$19\,au) with a major axis position angle with respect to the North of 44\arcdeg. We achieve a 1$\sigma$ sensitivity of $\sim$$2.3\times10^{-17}$\,W\,m$^{-2}$\,Hz$^{-1}$\,sr$^{-1}$ ($\sim$70\,mJy\,beam$^{-1}$) in the individual channels. We perform photometry by considering a rectangular aperture extending $\pm$115\,au in the horizontal direction (measured from the stellar position, see Figure \ref{fig:mom0_xprofile}) and $\pm$30\,au in the vertical direction (measured from the mid-plane). This yields a total flux of $(1.6\pm0.2)\times10^{-19}$\,W\,m$^{-2}$ ($9.8\pm1.4$\,Jy\,km\,s$^{-1}$), where the error is random without any systematic calibration uncertainty taken into account. We did not correct for the primary beam, as it changes the total flux only within the quoted error bars (the same applies for the continuum, section \ref{sec:continuum}). The measured flux is consistent with the value of $(1.7\pm0.4)\times10^{-19}$\,W\,m$^{-2}$ ($10.3\pm2.3$\,Jy\,km\,s$^{-1}$) derived from single dish observations by \citet{Higuchi_etal_2017}. The same method yields $3\sigma$ upper limits of $5.9\times10^{-20}$\,W\,m$^{-2}$ (3.6\,Jy\,km\,s$^{-1}$) for CS~10--9 and $2.9\times10^{-20}$\,W\,m$^{-2}$ (1.8\,Jy\,km\,s$^{-1}$) for SiO~11-10.

Figure \ref{fig:mom0_xprofile} (right) shows the emission profile along the x-axis of the moment-0 map, obtained by integrating within $\pm$30\,au in the vertical direction. Both the moment-0 map and the emission profile are suggestive of an asymmetry with a peak on the SW side of the disk. The same asymmetry is seen for the CO emission \citep{Dent_etal_2014,Matra_etal_2017} and tentatively also for \ion{C}{2} \citep{Cataldi_etal_2014}. However, the SW/NE flux ratio within the rectangular box of Figure \ref{fig:mom0_xprofile} is not significant at $0.9\pm0.2$ (we calculate the error on the ratio $\frac{a}{b}$ by propagating the error like this: $\sigma^2_{a/b}=\frac{1}{b^2}\sigma^2_a + \frac{a^2}{b^4}\sigma^2_b$). The SW/NE ratio of the peaks in the emission profile is $1.3\pm0.5$. Thus, there is not necessarily more flux on the SW side of the disk, but the flux seems to be more compact.

\begin{figure}
\plotone{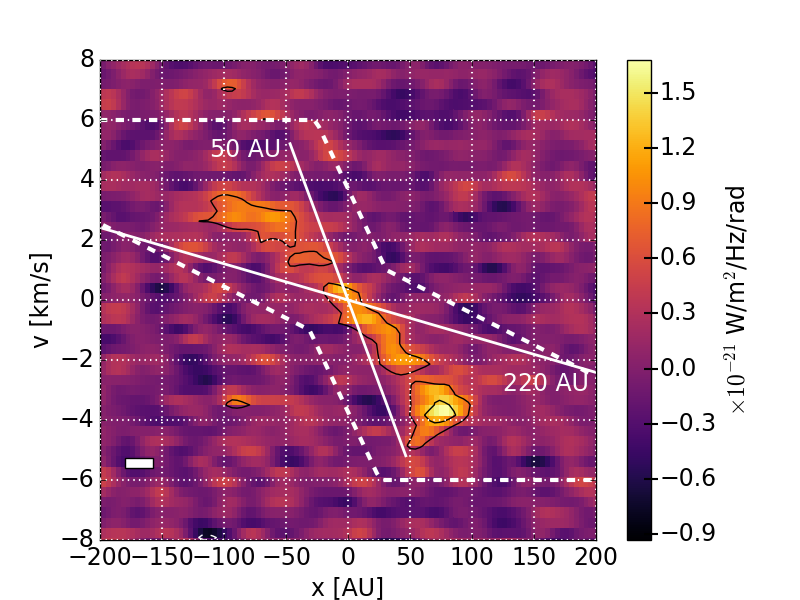}
\caption{Position-velocity diagram of the \ion{C}{1} emission. Contours are drawn at 3$\sigma$ intervals. The spectro-spatial resolution is illustrated in the lower left by the white rectangle. The two solid lines show the radial velocity (for circular Keplerian orbits around a 1.75\,M$_\Sun$ star seen edge-on) at 50\,au and 220\,au, the approximate radial extent of the CO \citep{Matra_etal_2017}. The dashed lines show the region in x-v-space used to derive a moment 0 map and emission profile with improved SNR.\label{fig:pv_diagram}}
\end{figure}

Figure \ref{fig:pv_diagram} shows the position-velocity (pv) diagram (i.e.\ the data cube integrated along the vertical spatial direction within $\pm 30$\,au). This figure thus shows the radial velocity of the emission as a function of the projected position along the disk mid-plane. By using the pv diagram, we can constrain the radial distribution of the emission despite the edge-on orientation of the disk. Indeed, figure \ref{fig:pv_diagram} also shows the radial velocity for circular Keplerian orbits at 50\,au and 220\,au around a 1.75\,M$_\Sun$ star \citep{Crifo_etal_1997}, seen edge-on. This is the approximate radial extent of the CO as found by \citet{Matra_etal_2017}. As can be seen, no significant \ion{C}{1} emission is detected beyond these lines, suggesting that most \ion{C}{1} emission is confined to the same region as the CO.

\begin{figure}
\plottwo{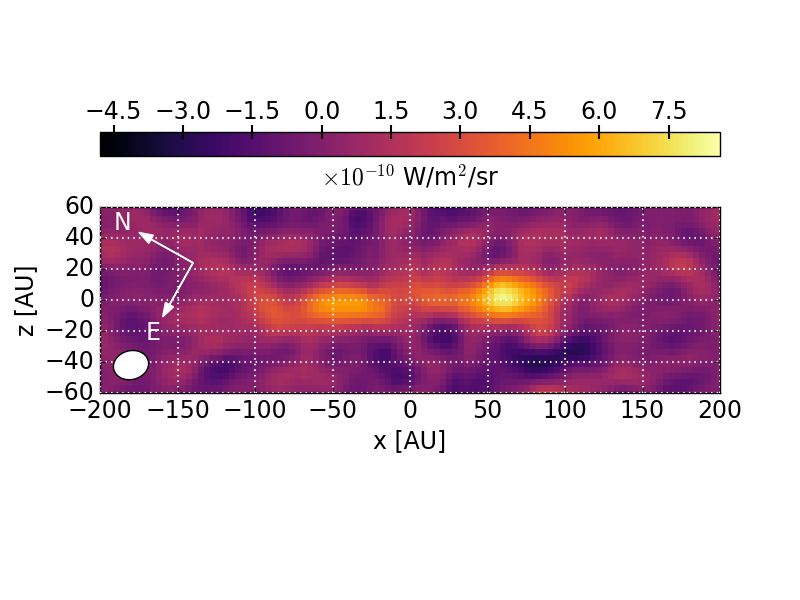}{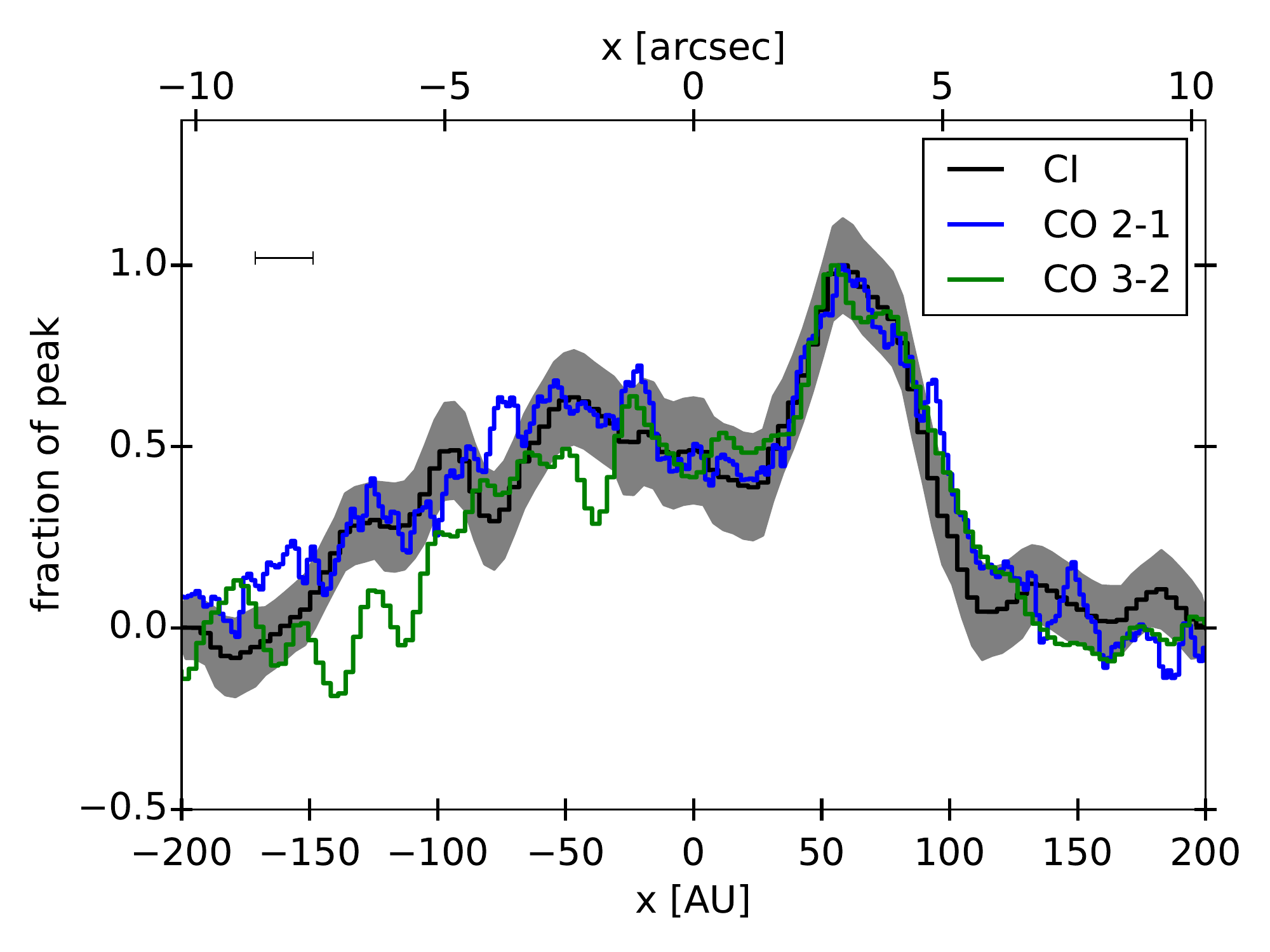}
\caption{Same as Figure \ref{fig:mom0_xprofile}, but using only the region in p-v space illustrated by the dashed lines in Figure \ref{fig:pv_diagram} to improve the SNR. For the moment 0 (left), no contours are shown since by construction, the noise levels are now non-uniform and depend on $x$. The color scale is the same as in Figure \ref{fig:mom0_xprofile}. In the emission profile (right) we also included the CO 3--2 and 2--1 emission as observed by \citet{Matra_etal_2017}. \label{fig:mom0_xprofile_pv_restriction}}
\end{figure}

We may use the pv diagram to define a region in p-v space that contains all significant emission. Then, when integrating over the spectral axis, we only take data points within this region into account (rather than everything within $\pm6$\,km\,s$^{-1}$ as was done to produce Figure \ref{fig:mom0_xprofile}). The region is illustrated by the dashed lines in Figure \ref{fig:pv_diagram} and the resulting moment 0 map and emission profile are shown in Figure \ref{fig:mom0_xprofile_pv_restriction}. The SW/NE flux ratio (within the same box as in figure \ref{fig:mom0_xprofile}) is now $1.2\pm0.2$. Most importantly, the SNR of the emission profile is significantly improved and the SW/NE asymmetry more clearly visible. We measure a SW/NE peak ratio of $1.6\pm0.4$. Thus, the significance of the peak ratio asymmetry is only marginal. However, in Figure \ref{fig:mom0_xprofile_pv_restriction} (right) we also show the profiles of the CO~3--2 and 2--1 emission \citep[see Figure 2 of][]{Matra_etal_2017}, for which the SW/NE flux ratios are $1.08\pm0.08$ (CO~2--1) and $1.49\pm0.13$ (CO~3--2) and the peak ratios are $1.42\pm0.14$ (CO~2--1) and $1.51\pm0.14$ (CO~3--2). The \ion{C}{1} emission follows the CO~2--1 emission surprisingly well, with the same distinct peak on the SW side of the disk and the same asymmetric extent (out to $\sim$150\,au in the NE and $\sim$100\,au in the SW). This suggest that the asymmetry is indeed real. This is a surprising result. The asymmetry in CO can be readily understood from its short (less than one orbit) lifetime due to photodissociation: if CO is primarily produced in a clump, photodissociation prevents CO from spreading azimuthally, thus preserving the asymmetry. On the other hand, the C produced from CO photodissociation is expected to spread in azimuth within a few orbits. In section \ref{sec:C_asymmetry_origin} we discuss possible solutions to this puzzle.

From the moment 0 maps, it is also apparent that the observed emission is slightly tilted, with the NE side below and the SW side above the midplane of the main outer disk (defined by $z=0$). A similar tilt is seen for CO \citep{Dent_etal_2014,Matra_etal_2017}. As is discussed by \citet{Matra_etal_2017}, two reasons might be imagined for this observation: either the gas disk is indeed tilted with respect to the dust disk, or the tilt is a projection effect, resulting from an azimuthally asymmetric gas disk in combination with a slight deviation from a perfect edge-on inclination. Interestingly, the inner dust disk seen in scattered light has a similar tilt \citep[e.g.][]{Milli_etal_2014,Apai_etal_2015,Millar-Blanchaer_etal_2015}. This dust component known as \emph{warp} or \emph{secondary disk} is localised inwards of 80\,AU \citep{Lagrange_etal_2012} and seems thus slightly inwards of the gas. Also, it has already been suggested that this inner dust disk is not perfectly edge-on \citep[e.g.][]{Milli_etal_2014,Millar-Blanchaer_etal_2015}.

The reader interested in the procedures to estimate the errors quoted in this section is referred to Appendix \ref{sec:error_calculation_details}.

\subsection{Deprojection of the \ion{C}{1} emission}
\begin{figure}
\plotone{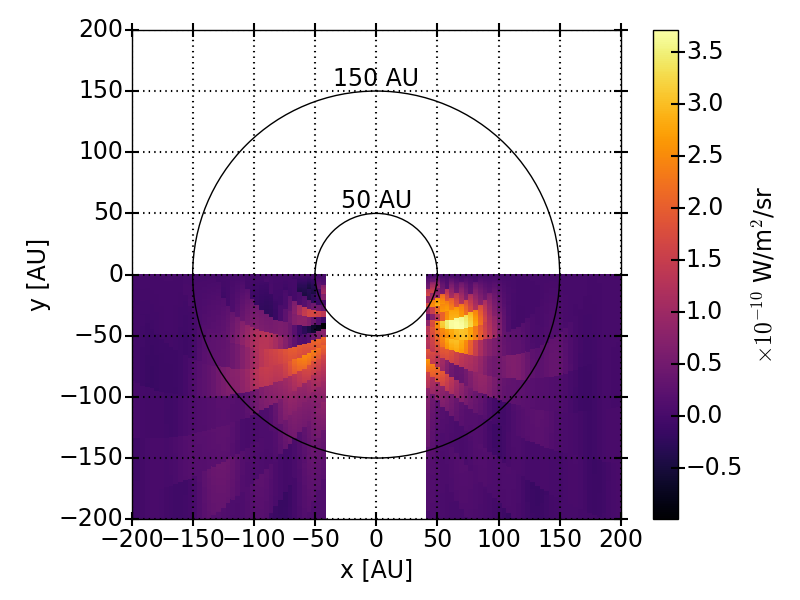}
\caption{Deprojection (face-on view) of the \ion{C}{1} emission derived from the pv diagram (figure \ref{fig:pv_diagram}), where we chose to place all flux in front of the sky plane (the line of sight runs in the positive direction of the $y$-axis). Points with $|x|<40$\,au are masked.\label{fig:deprojection}}
\end{figure}

Assuming that the gas follows circular Keplerian orbits, we can use the pv diagram (figure \ref{fig:pv_diagram}) to obtain a face-on view of the emission \citep[e.g.][]{Dent_etal_2014,Matra_etal_2017}. Indeed, each point in the pv diagram corresponds to two points in the xy-plane of the disk (where we define $y$ as the coordinate running along the line of sight), one in front and one behind the sky plane. There remains the degeneracy of how to distribute the flux of a given pv point among the two points in the xy-plane. As discussed by \citet{Matra_etal_2017}, the degeneracy can be broken if the disk is not perfectly edge-on. However, for simplicity, we follow \citet{Dent_etal_2014} and assume that the disk is edge-on. Our primary interest is to illustrate the radial extent of the emission and the position of the clump. We thus choose to place all flux in front of the sky plane, but other physically motivated choices are possible \citep{Dent_etal_2014}. Figure \ref{fig:deprojection} shows the deprojection. Points of the pv diagram with $|x|<40$\,au were masked because the radial velocity in this regions tends towards zero for all radii, i.e.\ it becomes difficult to assign a radius to a given radial velocity. Emission is seen approximately out to 150\,au. The clump appears at a similar position angle as seen in CO \citep{Dent_etal_2014}. Details of the deprojection procedure are described in appendix \ref{sec:deprojection_details}. Since the optical depth of the emission is not negligible, the deprojection does not show the true distribution of the emission in the $xy$-plane but rather how much emission the observer receives from various locations in the $xy$-plane.

\subsection{Continuum}\label{sec:continuum}
\begin{figure}
\plotone{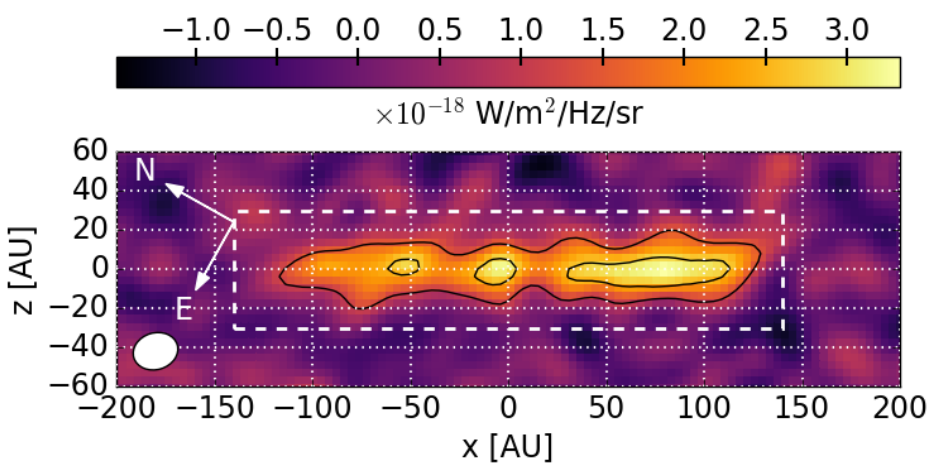}
\caption{ALMA continuum image of the $\beta$~Pic disk at 480\,GHz. Contours are drawn at intervals of 3$\sigma$. The white dashed rectangle indicates the region within which the flux was measured.\label{fig:continuum}}
\end{figure}

The continuum image at $\sim$485\,GHz is shown in figure \ref{fig:continuum}. The 1$\sigma$ noise level is $4.4\times10^{-19}$\,W\,m$^{-2}$\,Hz$^{-1}$\,sr$^{-1}$ (1.3\,mJy\,beam$^{-1}$). The beam size is 1.19\arcsec$\times$0.96\arcsec\ (23\,au$\times$19\,au) with the position angle of the major axis being 46\arcdeg. We measure a total flux of $(1.12\pm0.07)\times10^{-27}$\,W\,m$^{-2}$\,Hz$^{-1}$ ($112\pm7$\,mJy) in the rectangular region ($\pm140$\,au from the star along $x$ and $\pm$30\,au from the mid-plane along $z$) indicated in the figure (see appendix \ref{sec:error_calculation_details} for details on the error calculation). The measured flux is consistent with a Rayleigh-Jeans extrapolation of the flux measured by ALMA at 870\,$\mu$m \citep{Dent_etal_2014}, and is a factor of $\sim$2 below of what is expected from the infrared to mm SED fit by \citet{Vandenbussche_etal_2010}.

As was already seen in the ALMA data by \citet{Dent_etal_2014}, the continuum is brighter on the SW side of the disk at projected separations between 50--100\,au. However, the SW/NE integrated flux ratio is only $1.17\pm0.14$. In any case, the asymmetry is analogous to what is seen in thermal mid-IR images between 8.7 and 18.3\,$\mu$m \citep{Telesco_etal_2005} as well as for C and CO.

\section{Modelling}\label{sec:modelling}
\subsection{Simple estimation of the total carbon mass}\label{sec:simple_mass_estimation}
In this section, we estimate the total carbon mass in the $\beta$~Pic disk under some simplifying assumptions. With a 1D model, we want to reproduce the \ion{C}{1} flux measured in the present work and the \ion{C}{2} flux measured with \textit{Herschel}/PACS \citep{Pilbratt_etal_2010,Poglitsch_etal_2010} by \citet[][OBSID 1342198171]{Brandeker_etal_2016} (we sum the flux values of the PACS spaxels listed in their table 1). \textit{Herschel}/HIFI also measured the \ion{C}{2} flux \citep{Cataldi_etal_2014}. However, the flux measured by PACS is likely a better estimate of the total \ion{C}{2} flux from the disk, because the HIFI half power beam width is only $\sim$200\,au.

We need to compute the statistical equilibrium of the level populations, as in general the emission cannot be assumed to be in local thermal equilibrium (LTE). The gas in the $\beta$~Pic disk is thought to be of secondary origin and thus depleted in hydrogen \citep[e.g.][]{Matra_etal_2017}. We thus assume that the dominant collider species is electrons, and that photoionisation of carbon is the main electron source \citep{Fernandez_etal_2006,Kral_etal_2016}, i.e.\ the density of ionised carbon equals the electron density. Under these assumptions, the total carbon mass is estimated in two steps. For a given kinetic temperature, we first determine the amount of ionised carbon necessary to reproduce the \ion{C}{2} emission. The second step is to determine the amount of neutral carbon necessary to reproduce the \ion{C}{1} flux observed by ALMA, using the electron density derived in the first step.

To solve the statistical equilibrium and radiative transfer, we use \texttt{pythonradex}\footnote{\url{https://github.com/gica3618/pythonradex}}, a python implementation of the \textsc{RADEX} code \citep{vanderTak_etal_2007} with atomic data from the LAMDA database\footnote{\url{http://home.strw.leidenuniv.nl/~moldata/}} \citep{Schoier_etal_2005}. This code uses an escape probability formalism to solve the radiative transfer and assumes the geometry of a uniform sphere. In general, the radiative transfer depends on the geometry of the emitting region. The gas observed around $\beta$~Pic is clearly non-spherical and non-symmetric. The assumption of a uniform sphere is thus a clear limitation of this model. The masses derived here should therefore be considered first order estimates.

We choose the size of the sphere such that its volume corresponds to the volume of an elliptic torus with semi-major axis of 35\,au in the radial direction and semi-minor axis of 10\,au in the vertical direction (see Figures \ref{fig:mom0_xprofile},\ref{fig:pv_diagram} and \ref{fig:mom0_xprofile_pv_restriction}). Note that even in the optically thin case, such an assumption on the scale over which the emission is produced is necessary, unless the emission is in LTE. This is because for a given mass, the electron density depends on the assumed volume.

We include radiative excitation and de-excitation (hereafter simply (de-)excitation) by the CMB, the stellar radiation (at 85\,au) and the dust continuum, where the latter is taken at 85\,au as seen in figure B1 of \citet{Kral_etal_2017} and is the dominant component. However, it turns out that including radiative (de-)excitation from these sources does not change our results because (de-)excitation is dominated by collisions and spontaneous emission.

We then consider a wide range of kinetic temperatures $T_\mathrm{kin}$ from 40\,K to 1000\,K. For lower temperatures, the \ion{C}{2} line quickly becomes strongly optically thick such that meaningful constraints on the mass are no longer possible (i.e. the model cannot reproduce the observed flux regardless of how much the mass is increased). However, based on our more detailed modelling in section \ref{sec:3D_modelling}, we deem lower temperatures unlikely. Detailed thermodynamical modelling by \citet{Kral_etal_2016} also suggests that the temperature is above $\sim$50\,K within 100\,au, although in their model, the temperature drops to 20\,K at 150\,au.

\begin{figure}
\plotone{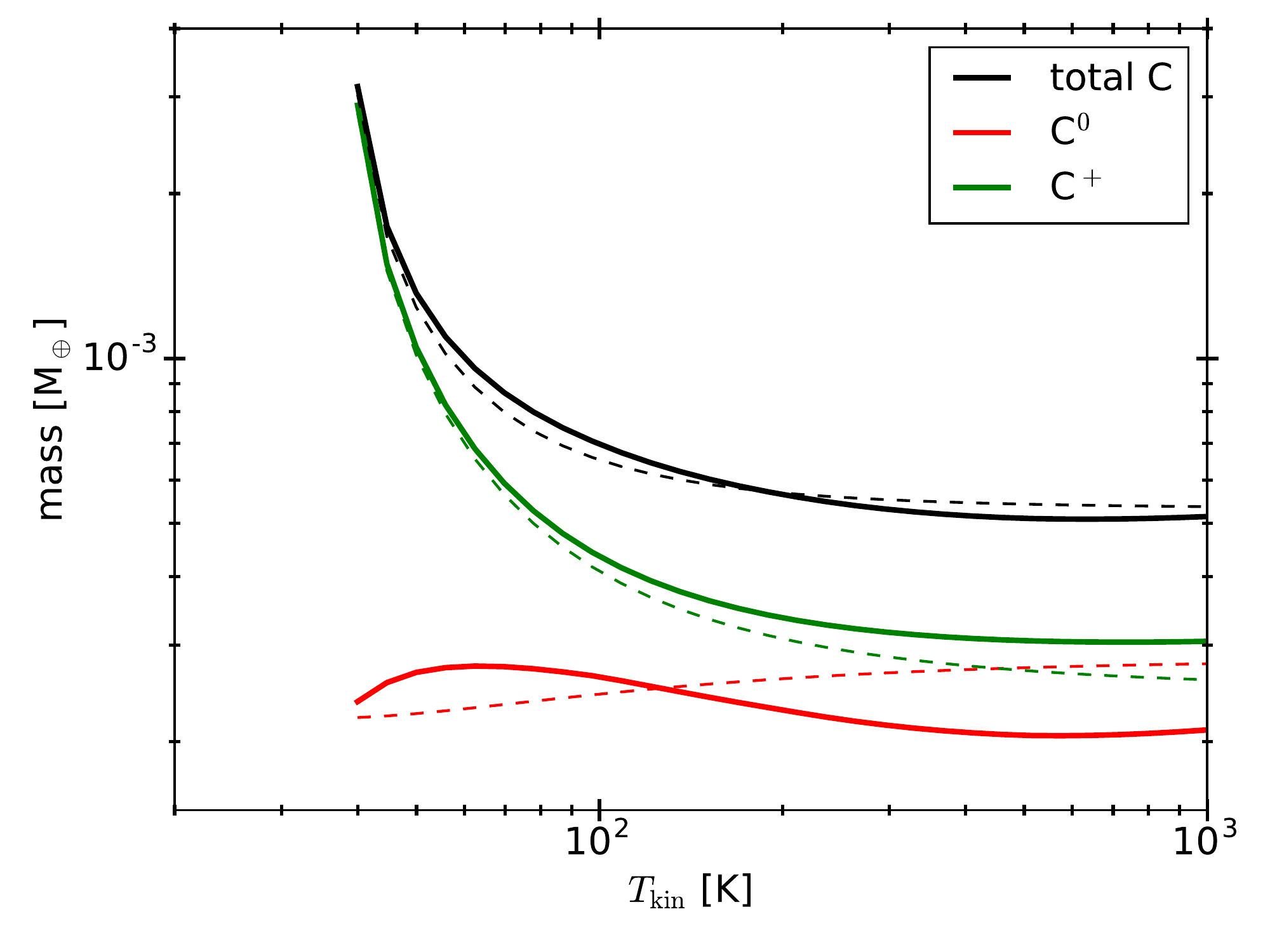}
\caption{Mass of C$^0$, C$^+$ and total C (derived from the model described in section \ref{sec:simple_mass_estimation}) as a function of the assumed kinetic temperature. For the full lines, the statistical equilibrium has been solved, while for the dashed lines, LTE has been assumed.\label{fig:C_mass}}
\end{figure}

Figure \ref{fig:C_mass} shows the derived total carbon mass as a function of $T_\mathrm{kin}$. The figure also shows the individual C$^0$ and C$^+$ masses. To assess the importance of non-LTE effects, we also show corresponding curves for which LTE has been assumed. Both \ion{C}{1} and \ion{C}{2} are generally speaking close to LTE. For higher temperatures, the C$^0$ mass required to reproduce the observed \ion{C}{1} flux is \emph{higher} in LTE than in non-LTE. This simply happens because in LTE, higher levels get populated more quickly with increasing temperature, thus de-populating the level that produces the \ion{C}{1} emission. The maximum optical depths encountered for the considered temperature range are 0.5 for \ion{C}{1} and 3.9 for \ion{C}{2}.

From Figure \ref{fig:C_mass}, we determine a total carbon mass between $5\times10^{-4}$ and $3.1\times10^{-3}$\,M$_\earth$. The lower bound is quite robust to changes of the size of the emitting region as it is close to the LTE value. For example, increasing or decreasing the radius of the emitting sphere by 50\% does not change the lower bound by more than 15\%. On the other hand, it is clear that the upper bound is more uncertain as it can quickly increase if one allows for lower temperatures and/or smaller emitting volumes (i.e.\ increased optical depth). Another parameter that can strongly affect the optical depth (and thus the upper bound of the mass range) is the assumed line broadening parameter. Here we used $b = 0.57$\,km\,s$^{-1}$, derived in appendix \ref{sec:line_broadening}.

Previous studies estimating the carbon gas mass in the $\beta$~Pic disk had only the \ion{C}{2} flux and line profile at disposal. These more detailed models derived higher masses by using the spectrally resolved \ion{C}{2} observations from Herschel/HIFI: \citet{Cataldi_etal_2014} obtained $1.3\times10^{-2}$\,M$_\earth$ while \citet{Kral_etal_2016} derived $1.5\times10^{-2}$\,M$_\earth$ (the latter study also used an upper limit on the \ion{C}{1} flux). In the \citet{Kral_etal_2016} model, the total C mass is dominated by ionised carbon that is located beyond $\sim$100\,au. The \ion{C}{1} flux is of little importance for the total C mass budget of that model. The discrepancy can thus not be explained by the fact that we include the \ion{C}{1} measurement. On the other hand, most of the carbon in the \citet{Cataldi_etal_2014} model is located between 150 and 300\,au with an ionisation fraction of roughly 50\%, i.e.\ there is a significant contribution of neutral carbon to the total mass. However, our ALMA data show no \ion{C}{1} emission beyond $\sim$120\,au. That model indeed over-predicted the \ion{C}{1} flux by a factor of $\sim$20, suggesting that it contains too much neutral carbon, partially explaining the difference in the mass estimates. But \citet{Cataldi_etal_2014} also derived a higher mass of ionised carbon compared to our estimate. In addition, their fit to the resolved \ion{C}{2} line profile (see their figure 2b) clearly suggests that \ion{C}{2} emission beyond 150\,au is needed to fit the line core (this is also an issue for our 3D models (section \ref{sec:CII_line_core_issue}) that generally do a bad job in fitting the \ion{C}{2} line core). So maybe there is largely ionised carbon gas beyond $\sim$150\,au present and the reason why we derive a lower ionised carbon mass here is because we assumed (based on the \ion{C}{1} data) a too small volume (i.e.\ too high density and thus more excitation, and therefore less mass needed). However, even when increasing the volume of our 1D model, we still derive a lower mass of ionised carbon compared to \citet{Kral_etal_2016} and \citet{Cataldi_etal_2014} and can thus not fully resolve the discrepancy, which might be due to the different assumptions of the models. The relatively small HIFI beam (FWHM of $\sim$200\,au) could also play a role. For example, deviations from axisymmetry in the distribution of ionised carbon could introduce additional uncertainty in the mass estimates of \cite{Cataldi_etal_2014} and \citet{Kral_etal_2016} since \ion{C}{2} emission from large radii is only detected by HIFI if it arises close to the line of sight towards the star.

\subsection{3D modelling}\label{sec:3D_modelling}
In this section, we model the 3D spatial distribution of the carbon gas using our new ALMA observations of \ion{C}{1} and the previously published spectrally resolved \textit{Herschel}/HIFI observation of \ion{C}{2} \citep[][OBSID 1342238190]{Cataldi_etal_2014}. For a given, arbitrary distribution of carbon gas, we first solve the ionisation balance in each grid cell. We use the ionisation rate from \citet{Zagorovsky_etal_2010} (scaled with distance to the star) and assume that the gas is optically thin to ionising photons. The star is the main source of ionising photons, but ionisation from the interstellar radiation field (ISRF) is also included. The ionisation balance is solved analytically by assuming that all electrons come from the photoionisation of carbon. The ionisation fraction thus only depends on the distance to the star, the local gas density and the kinetic temperature (via the recombination coefficient). Next, we use the derived electron density to solve the statistical equilibrium (SE) of the level populations using atomic data from the LAMDA database. The following processes can (de)excite an atom: spontaneous emission, collisions with electrons (we neglect other colliders) and radiative (de)excitation by line photons and the background radiation field, where the latter is assumed to be composed of the CMB, the star and the dust continuum. For the dust continuum, we employ the field shown in figure B1 of \citet{Kral_etal_2017} (that figure actually shows the field in the midplane of the disk at the position angle of the clump (L.\ Matr\`a, private communication), but for simplicity, we only take the radial dependence into account). In principle, a full radiative transfer computation is necessary to solve the SE. However, we take a simplified approach and assume that the line emission is essentially optically thin. Basically, while the emission can become optically thick along the disk mid-plane (in our best-fit models, the maximum optical depth is typically of the order of 1), it is optically thin in the vertical direction, i.e.\ the photon escape fraction is high. Thus, we assume that the background field is not attenuated by the gas (all atoms are subject to the full background field) and that (de)excitation by line photons can be neglected. These assumptions make the SE easy to solve. We further discuss this approximation in appendix \ref{sec:approx_SE}. The background radiation turns out to be unimportant for our models.

Having solved the SE, we compute the emitted spectrum for each grid cell, red- or blue-shifting it according to its radial velocity. The final step is then to ray-trace the emission along the line of sight to take optical depth into account. The result is a model data cube that can be compared to observations. For simplicity, we consider isothermal models and fix $T_\mathrm{kin}=75$\,K everywhere. We found that for lower temperatures, the \ion{C}{1}/\ion{C}{2} flux ratio tends to be too high, while the inverse is true for higher temperatures.

To compare to the \ion{C}{2} data, we take the corresponding model data cube, multiply by the HIFI beam and integrate spatially to get a model HIFI spectrum \citep{Cataldi_etal_2014}. For the \ion{C}{1} ALMA observations, we convolve the model data cube to the same spatial and spectral resolution as the observations and multiply by the primary beam. The residual moment 0 maps (respectively pv diagrams) shown in figures \ref{fig:ring}, \ref{fig:ring_clump} and \ref{fig:eccentric} are obtained by subtracting the model moment 0 map (pv diagram) from the data moment 0 map (pv diagram) shown in figure \ref{fig:mom0_xprofile} (figure \ref{fig:pv_diagram}).

\subsubsection{Uniform ring model}\label{sec:ring}
We first consider a simple, symmetric model consisting of a ring with uniform surface density. The number density reads
\begin{equation}
n_\mathrm{ring}(r,z) = 
\begin{cases}
\frac{\Sigma}{\sqrt{2\pi}H(r)} \cdot \exp\left( -\frac{z^2}{2(H(r))^2}\right) & \text{if } r_\mathrm{min}\leq r \leq r_\mathrm{max}\\
0  & \text{otherwise}
\end{cases}
\end{equation}
where $r$ and $z$ are cylindrical coordinates (with $z$ perpendicular to the disk mid-plane), $H(r)$ is the scale height, $r_\mathrm{min}$ and $r_\mathrm{max}$ define the radial extent of the ring and $\Sigma$ is the constant surface density. The scale height in hydrostatic equilibrium for a vertically isothermal disk is given by \citep[e.g.][]{Armitage_2009}
\begin{equation}\label{eq:scale_height}
H(r) = \sqrt{\frac{kTr^3}{\mu m_p G M_*}}
\end{equation}
with $k$ the Boltzmann constant, $T$ the temperature, $m_p$ the proton mass, $G$ the gravitational constant and $M_*$ the stellar mass. For the mean molecular weigh $\mu$, we follow \citet{Matra_etal_2017} and assume $\mu=14$. At 85\,au and for $T=75$\,K, the scale height is 4.2\,au.

We compute a grid of models over the parameter space listed in the first four rows of table \ref{tab:parameter_space_ring_clump}. Note that we also test different values for the (dynamical) stellar mass that affects the orbital velocities (and scaleheight). To each model, we assign a $\chi^2$ measure by using expressions analogous to equation 2 in \citet{Booth_etal_2017} (we take the correlation of neighbouring data points into account by using an appropriate noise correlation ratio for each data set). We sum the $\chi^2$ from the \textit{Herschel}/HIFI data (\ion{C}{2}) and the ALMA data (\ion{C}{1}).

Figure \ref{fig:ring} shows the \ion{C}{1} residuals in the $xz$ and $xv$ plane as well as the \ion{C}{2} line emission of the model with the lowest $\chi^2$. The corresponding model parameters are given in table \ref{tab:parameter_space_ring_clump}. The model provides a decent fit to the data, but unsurprisingly is not able to reproduce the clump observed in the SW. Thus, we consider a more complicated model by adding a clump to the uniform ring in the next section. Such a model has no direct physical justification, but is useful to empirically constrain the spatial distribution of the gas and get an estimate of the gas mass.

\subsubsection{Uniform ring with a clump}\label{sec:ring_clump}
\begin{figure}
\plotone{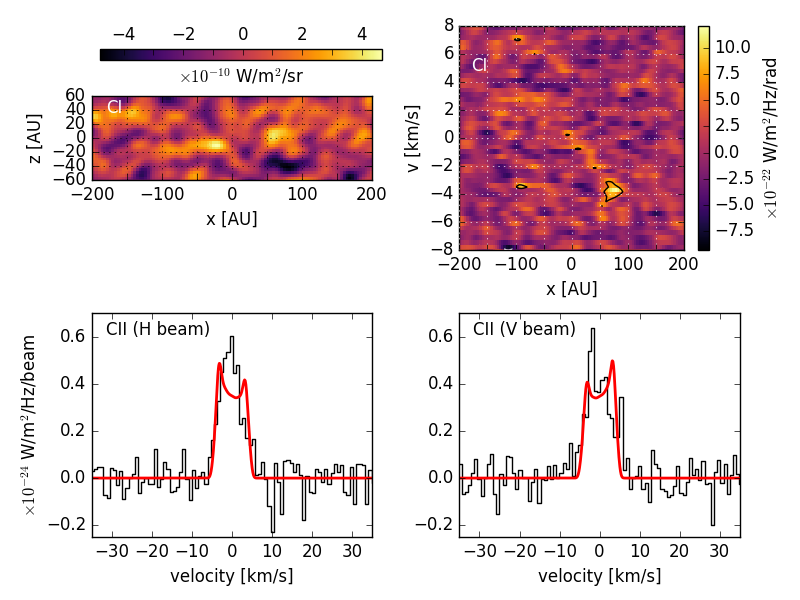}
\caption{Comparison of the uniform ring model to the data. The top row shows the residual moment 0 map (left) and pv diagram (right) for the \ion{C}{1} emission. Contours are in steps of 3$\sigma$. The bottom row compares the modeled \ion{C}{2} emission (red lines) to the HIFI data (black lines).\label{fig:ring}}
\end{figure}

\begin{figure}
\plotone{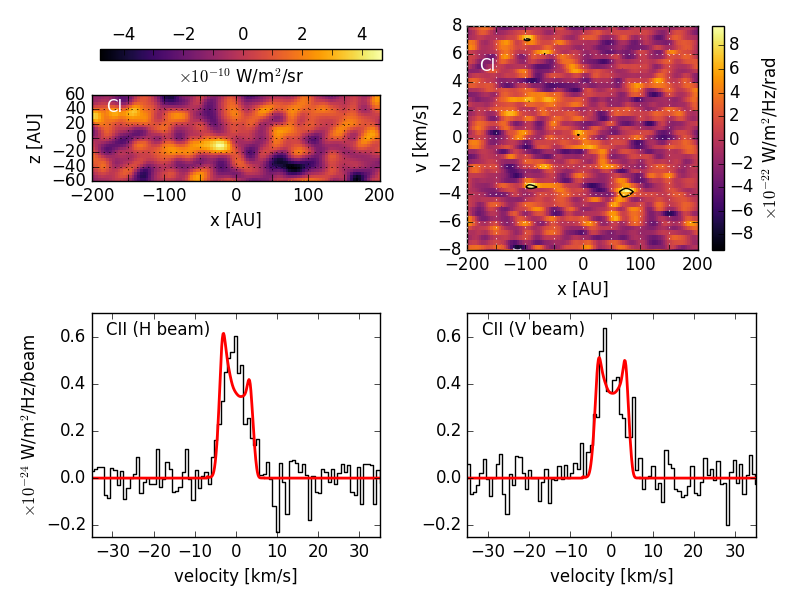}
\caption{Same as figure \ref{fig:ring}, but for the `ring + clump' model.\label{fig:ring_clump}}
\end{figure}

The clump is modelled as
\begin{equation}
n_\mathrm{clump}(x,y,z) = n_0 \cdot \exp\left( -\frac{(x-x_0)^2+(y-y_0)^2}{2\sigma_\mathrm{xy}^2}\right) \cdot \exp\left( -\frac{z^2}{2(H(r))^2}\right)
\end{equation}
where $x$ runs along the disk mid-plane in the sky plane and $y$ along the line of sight (and $x^2+y^2=r^2$). Furthermore, $x_0=r_\mathrm{c}\cos(\phi_{\mathrm{c}})$ and $y_0=r_\mathrm{c}\sin(\phi_{\mathrm{c}})$ designate the center of the clump, where we have defined $r_\mathrm{c}$ as the clump's radial distance to the star and $\phi_{\mathrm{c}}$ as its azimuthal angle in the x-y-plane. We fix $\phi_{\mathrm{c}}=-32\arcdeg$ \citep{Matra_etal_2017}. The standard deviation of the clump density distribution in the $x$-$y$-plane is $\sigma_\mathrm{xy}$ and the density at the centre $n_0$. The total carbon number density is then given by $n_\mathrm{C}=n_\mathrm{ring}+n_\mathrm{clump}$. We compute models over the parameter space listed in table \ref{tab:parameter_space_ring_clump}. Figure \ref{fig:ring_clump} is the analogue of figure \ref{fig:ring} for the best `ring + clump' model. As can be seen, the model does a slightly better job in modelling the clump.

\begin{table}
\renewcommand{\thetable}{\arabic{table}}
\centering
\caption{Explored parameter space and best fit values for the `ring' and `ring + clump' models fitting the \ion{C}{1} ALMA data and the \ion{C}{2} \textit{Herschel}/HIFI data simultaneously. The number of values explored for each parameter is denoted by $n$. We also indicate whether the values are uniformly distributed in linear or logarithmic space. $M_\mathrm{ring}$ and $M_\mathrm{clump}$ are the masses of the ring and the clump respectively. The reference mass of $\beta$~Pic is assumed $M_\mathrm{\beta Pic}=1.75$\,$M_*$ \citep{Crifo_etal_1997}. We also list the constant surface density of the ring $\Sigma$, the mid-plane number density of the ring $n_\mathrm{mid}$ at 85\,au and, for the `ring + clump' model, the number density at the centre of the clump $n_\mathrm{clump}$ resulting from the superposition of the ring and the clump.} \label{tab:parameter_space_ring_clump}
\begin{tabular}{cccccccc}
\tablewidth{0pt}
\hline
\hline
parameter & unit & min & max & $n$ & spacing & \multicolumn{2}{c}{best fit} \\
&&&&&&`ring'&`ring + clump'\\
\hline
$M_*$ & $M_\mathrm{\beta Pic}$ &  0.6  & 1  & 3 & lin & 0.8 & 0.8 \\
$M_\mathrm{ring}$ & $M_\earth$ & $2.7\times10^{-4}$ & $2.7\times10^{-3}$ & 6 & log & $6.7\times10^{-4}$ & $6.7\times10^{-4}$\\
$r_\mathrm{min}$& au & 30 & 70 & 3 & lin & 50 & 50\\
$r_\mathrm{max}$& au & 120 & 160 & 3 & lin & 120 & 120\\
$M_\mathrm{clump}$ & $M_\earth$ & $2.7\times10^{-5}$ & $2.7\times10^{-4}$ & 6 & log & & $1.1\times10^{-4}$\\
$r_\mathrm{c}$ & au & 70 & 100 & 3 & lin && 70\\
$\sigma_\mathrm{xy}$& au & 20 & 40 & 3 & lin && 40\\
\hline
$\Sigma$& cm$^{-2}$ &&&&&$2.4\times10^{16}$ & $2.4\times10^{16}$ \\
$n_\mathrm{mid}(r=85\mathrm{au})$& cm$^{-3}$ &&&& & 140 & 140 \\
$n_\mathrm{clump}$& cm$^{-3}$ &&&&& & 280 \\
\hline
\end{tabular}
\end{table}

\subsubsection{Eccentric gas distribution}\label{sec:eccentric_gas_distribution}
The `ring + clump' model of the previous section is purely empirical and does not have a direct physical justification. As mentioned earlier, proposed mechanisms to get such a morphology are either a giant collision or resonant trapping of cometary bodies by a migrating giant planet. However, as we discuss in section \ref{sec:C_asymmetry_origin}, based on our new \ion{C}{1} data, we deem both of these possibilities unlikely.

Another way to get a morphology with an emission clump on one side of the disk is to relax the assumption of circular orbits and instead consider gas on eccentric orbits. In section \ref{sec:tidal_disruption}, we discuss how such orbits could arise in the first place. As an example, we here consider a model where the eccentricity of the orbits is uniformly distributed between two values $e_\mathrm{min}$ and $e_\mathrm{max}$ and where all orbits share the same pericentre and the same argument of periapsis. The pericentre is then a region of higher density and can mimic a clump as seen in the observations.

\begin{figure}
\plotone{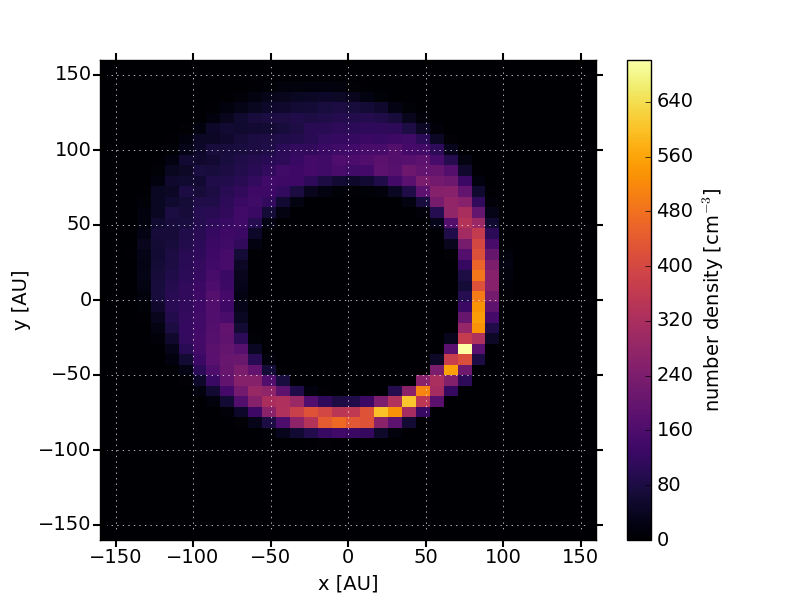}
\caption{Face-on view (mid-plane number density) of the C gas for the best fit model with eccentric orbits described in section \ref{sec:eccentric_gas_distribution}.\label{fig:eccentric_midplane_density}}
\end{figure}

\begin{figure}
\plotone{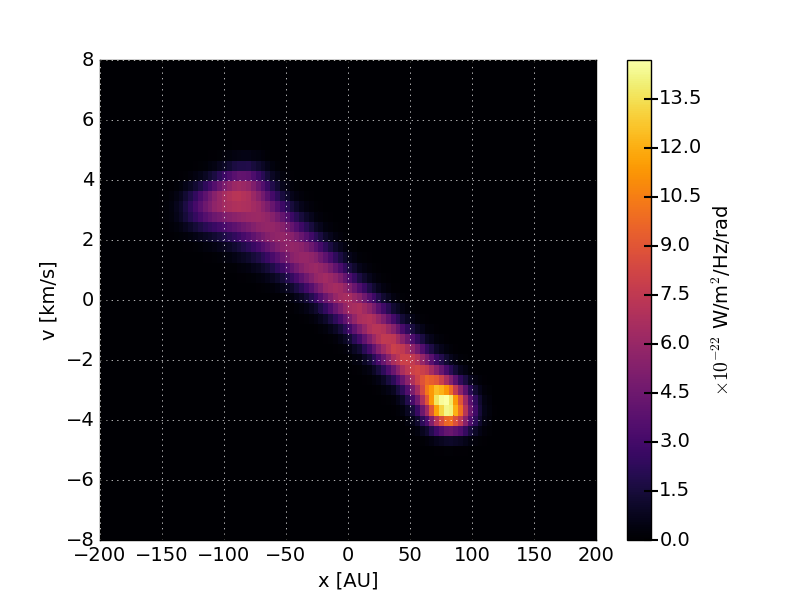}
\caption{PV diagram of the \ion{C}{1} emission for the best fit model with eccentric orbits, degraded to the spectro-spatial resolution of the ALMA data.\label{fig:eccentric_pv}}
\end{figure}

\begin{figure}
\plotone{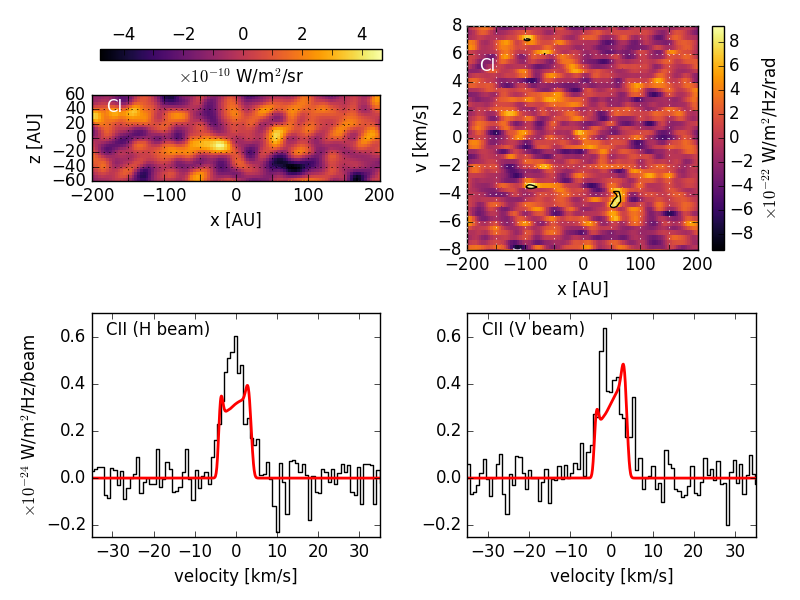}
\caption{Same as figure \ref{fig:ring}, but for the best fit model with eccentric orbits.\label{fig:eccentric}}
\end{figure}

The derivation of the gas density for this model is given in appendix \ref{sec:derivation_eccentric_orbits}. We again compute a grid of models with the parameters listed in table \ref{tab:parameter_space_eccentric_orbits}. Figure \ref{fig:eccentric_midplane_density} shows a face-on view of the mid-plane carbon gas density and figure \ref{fig:eccentric_pv} the pv diagram of the modelled \ion{C}{1} emission. As is seen in figure \ref{fig:eccentric}, this model is similarly effective in reproducing the observed \ion{C}{1} emission, although it has some difficulties to reproduce the \ion{C}{2} line. We emphasize that the purpose of this model is merely to demonstrate that eccentric gas orbits are able to reproduce the general morphology with the clump. A more detailed model will be presented in a forthcoming paper.

An interesting consequence of eccentric orbits is that the gas along the line of sight towards the star has non-zero radial velocity. Thus, emission (or absorption) towards the star is shifted with respect to the systemic velocity. For example, our best-fit model predicts that the emission peak towards the star appears 0.4\,km\,s$^{-1}$ blue-shifted. Another consequence is an additional velocity broadening compared to the circular case, with broadening parameter $b_\mathrm{ecc}\approx$0.4\,km\,s$^{-1}$ (this is smaller than and thus consistent with the broadening measured from the pv diagram, see Appendix \ref{sec:line_broadening}). The velocity shift and broadening depend on the eccentricity and orientation of the disk. To verify this velocity shift and constrain the model, we would need to know the absolute stellar radial velocity to better accuracy than $\sim$80\,m\,s$^{-1}$ (for a 5$\sigma$-detection of the shift).

The column densities of neutral carbon towards the star for the three models discussed in sections \ref{sec:ring}, \ref{sec:ring_clump} and \ref{sec:eccentric_gas_distribution} are (6--7)$\times10^{16}$\,cm$^{-2}$, slightly above the value of (2--4)$\times10^{16}$\,cm$^{-2}$ measured in absorption by \citet{Roberge_etal_2000}. For ionised carbon, the models range between $5\times10^{16}$\,cm$^{-2}$ and $1.2\times10^{17}$\,cm$^{-2}$, while \citet{Roberge_etal_2006} report $2\times10^{16}$\,cm$^{-2}$.

\begin{table}
\renewcommand{\thetable}{\arabic{table}}
\centering
\caption{Same as table \ref{tab:parameter_space_ring_clump}, but for the parameter space explored by the models with eccentric orbits. The total mass of the model is $M_\mathrm{tot}$, the pericentre distance is $r_\mathrm{per}$ and the argument of pericentre is $\omega$.} \label{tab:parameter_space_eccentric_orbits}
\begin{tabular}{ccccccc}
\tablewidth{0pt}
\hline
\hline
parameter & unit & min & max & $n$ & spacing & best fit\\
\hline
$M_*$ & M$_\mathrm{\beta Pic}$ &  0.6  & 1  & 3 & lin & 0.8 \\
$M_\mathrm{tot}$ & $M_\earth$ & $2.0\times10^{-4}$ & $3.3\times10^{-3}$ & 6 & log & $6.2\times10^{-4}$ \\
$e_\mathrm{min}$& - & 0 & 0.4 & 5 & lin & 0\\
$e_\mathrm{max}$& - & 0.1 & 0.5 & 5 & lin & 0.3\\
$r_\mathrm{per}$ & au & 60 & 100 & 3 & lin & 80\\
$\omega$& deg & -135 & 135 & 7 & lin & -45\\
\hline
\end{tabular}
\end{table}

\subsection{Absence of an atomic accretion disk}\label{sec:no_accretion_disk}
Recently, \citet{Kral_etal_2016} presented a model that computes the temperature, ionisation and hydrodynamical evolution of the atomic carbon and oxygen in the $\beta$~Pic disk. In this model, the atomic gas is produced in a parent belt from photodissociation of CO and then evolves viscously under the influence of the MRI, i.e.\ it forms an accretion and decretion disk. \citet{Kral_etal_2016} predict that the carbon is mostly neutral in the inner parts of the accretion disk.

\begin{figure}
\plotone{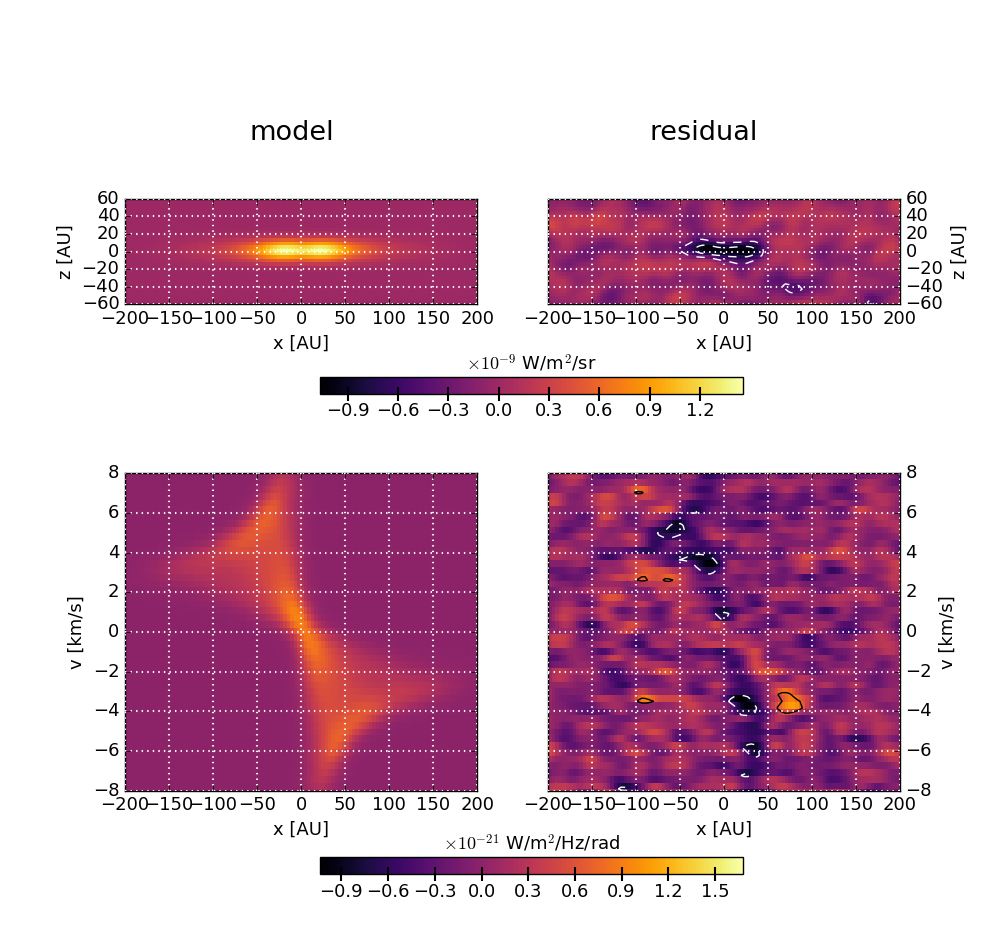}
\caption{Model \ion{C}{1} emission (left column) computed from the predictions by \citet{Kral_etal_2016}, and residuals (right column) when subtracting the model from the ALMA data. The top row show the moment 0, and the bottom row the pv diagram. Contours are in steps of 3$\sigma$, with negative contours drawn as dotted lines.\label{fig:Kral_model}}
\end{figure}

However, figures \ref{fig:pv_diagram} and \ref{fig:deprojection} indicate that no atomic accretion disk has formed (yet) as there seems to be little \ion{C}{1} emission inside of $\sim$50\,au (but see also section \ref{sec:discussion_accretion_disk}). To confirm this, we compute the \ion{C}{1} emission expected from the model and compare it to our data. We thus take the \citet{Kral_etal_2016} distribution of neutral carbon and electrons, temperature profile and scale height (see their figure 9) and compute the \ion{C}{1} emission with the methods described in section \ref{sec:3D_modelling}. Figure \ref{fig:Kral_model} shows the moment 0 map and pv-diagram of the model and the residuals when subtracting the model from the ALMA \ion{C}{1} data. It is clear from this figure that the accretion profile predicted by \citet{Kral_etal_2016} produces too much \ion{C}{1} emission in the inner regions of the disk. This is also clearly visible in the pv diagram where too much emission is present at high velocities.

\section{Discussion}\label{sec:discussion}
\subsection{Dynamical mass of the star}
For all the 3D models described in section \ref{sec:3D_modelling}, the grid searches indicate that a dynamical mass of $0.8M_*$ \citep[with $M_*=1.75M_\sun$ the assumed stellar mass,][]{Crifo_etal_1997} is preferred to reproduce the data. While this result has to be interpreted with care given that we did not derive any error bars, it is interesting to note that \citet{Olofsson_etal_2001} derived the same $0.8M_*$ dynamical mass by modelling spatially and spectrally resolved \ion{Na}{1} emission. They ascribed their finding to radiation pressure opposing the gravity of the star. In fact, radiation pressure on Na is so strong that a braking mechanism is required to keep it on the observed Keplerian orbits and preventing it from being blown out \citep[e.g.][]{Liseau_2003,Brandeker_etal_2004}. \cite{Fernandez_etal_2006} suggested that all species affected by radiation pressure are largely ionised and coupled into a single fluid by Coulomb interactions, with carbon acting as a braking agent. In this situation, neutrals are also expected to be coupled (the only exception would be neutrals that do not ionise yet still feel a significant radiation force). Thus, one might actually expect that \ion{Na}{1} and \ion{C}{1} share the same dynamics. The dynamical mass (respectively the radiation pressure coefficient) is coupled to the composition of the gas \citep{Fernandez_etal_2006} and could thus give interesting information on the elemental abundances. However, given the unknown uncertainty of the derived dynamical mass which could also be influenced by the assumptions of our models, we do not draw any further conclusions at this stage.
%A dynamical mass of $0.8M_*$ would correspond to an effective radiation force coefficient\footnote{See \citet{Fernandez_etal_2006} for the definition of $\beta_\mathrm{eff}$. For the gas to be self-braking, $\beta_\mathrm{eff}<0.5$} of $\beta_\mathrm{eff}=0.2$. The value of $\beta_\mathrm{eff}$ depends on the composition of the gas, with a more carbon-rich gas resulting in lower values because carbon does not feel any radiation pressure. Using equation 15 of \citet{Fernandez_etal_2006}, $\beta_\mathrm{eff}=0.2$ would correspond to a carbon overabundance of $\sim$45 compared to solar abundances of metals such as Na. 
%comment on effective beta implied by this (0.2) and why this is not compatible with the composition (see Alexis email)
%remember also that Na and CI have different spatial distributions
%why did Brandeker et al. (2004) not discuss this?

\subsection{Issue with the \ion{C}{2} line profile core}\label{sec:CII_line_core_issue}
All the 3D models presented in section \ref{sec:3D_modelling} have difficulties to reproduce the core of the \ion{C}{2} line profile (see figures \ref{fig:ring}, \ref{fig:ring_clump} and \ref{fig:eccentric}). The issue is particularly pronounced for the model with eccentric orbits. The fits to the \ion{C}{2} line profile by \citet{Cataldi_etal_2014} indicate that the line core requires ionised carbon beyond $\sim$150\,au to be present. As already discussed in section \ref{sec:simple_mass_estimation}, there might thus exist a gas component with a high ionisation fraction beyond $\sim$150\,au. This component might also be required to prevent the metals observed there from being blown out by radiation pressure (see section \ref{sec:metals_spatial_distribution}). It is possible that our simple models are unable to capture this extended gas component. For example, a more complex density and/or temperature profile might be required to reproduce it.  Note that while the \citet{Cataldi_etal_2014} model fits the \ion{C}{2} line profile, it would be a bad fit to our new \ion{C}{1} data as it strongly overpredicts the \ion{C}{1} flux.

\subsection{Overall timescale of CO and C production}\label{sec:C_production_timescale}
The strong spatial correlation between the \ion{C}{1} and CO emission (Figure \ref{fig:mom0_xprofile_pv_restriction}) suggests a scenario where the carbon is mainly produced from photodissociation of CO, i.e.\ the mass loss rate of CO equals the production rate of C. Thus, from the CO mass loss rate and our estimate of the total C mass, we can estimate the time since C (and CO) production started in the $\beta$~Pic disk, provided that no carbon has yet been removed.

\subsubsection{Revised CO lifetime}
It was previously thought that photodissociation of CO in the $\beta$~Pic disk is dominated by the ISRF. For example, taking self-shielding into account, \citet{Matra_etal_2017} calculated a CO lifetime of $\sim$300\,a in the clump against the ISRF. Here we revise this value, showing that photodissociation from the star actually dominates over the ISRF.

We compute the CO photodissociation rate using photodissociation cross sections from the Leiden Observatory database of `photodissociation and photoionisation of astrophysically relevant molecules'\footnote{\url{http://home.strw.leidenuniv.nl/~ewine/photo/}} \citep{Visser_etal_2009,Heays_etal_2017}. We calculate an unshielded lifetime against the ISRF \citep{Draine_1978,Lee_1984} of $\sim$130\,a. 

For the stellar spectrum, the basis is a PHOENIX model as described in \citet{Fernandez_etal_2006}. This model is then complemented with data in the UV from the Space Telescope Imaging Spectrograph (STIS) aboard the \textit{Hubble Space Telescope} (HST) \citep{Roberge_etal_2000} as well as from the the Far Ultraviolet Spectroscopic Explorer (FUSE) \citep{Bouret_etal_2002,Roberge_etal_2006}. This is important because $\beta$~Pic shows additional emission in the UV above the predictions from a standard stellar atmosphere model. These additional UV photons impact the calculated lifetime of CO. The lifetime corresponding to the observed stellar spectrum is $\sim$70\,a at 85\,au. However, since the light we observe has travelled through the full CO and C column, this is actually the lifetime of a CO molecule sitting behind this column. To obtain the unshielded lifetime against the star, we have to multiply by the CO self-shielding function \citep[table 6 in][]{Visser_etal_2009} and the shielding function of the C ionisation continuum \citep{Rollins_Rawlings_2012}, evaluated at the full CO and C$^0$ column densities against the star \citep{Roberge_etal_2000}. Thus, the unshielded lifetime against stellar photons at 85\,au is $\sim$20\,a.

From \citet{Matra_etal_2017} and \citet{Roberge_etal_2000}, we know the vertical column density at the clump location and the horizontal column density of CO against the star respectively. By applying a rough scaling based on the deprojected CO emission in \citet{Matra_etal_2017}, we estimate horizontal and vertical column densities at the clump location and along the line of sight to the star. For C$^0$, the horizontal column density against the star is taken from \citet{Roberge_etal_2000}, while for the vertical column density, we take our best-fit `ring + clump' model as a reference, which is also used to scale the \citet{Roberge_etal_2000} horizontal column density to the clump location. For the shielding towards the star, it is not difficult to see that the average CO molecule experiences half of the total column density against the star. For a Gaussian sphere, the average column density per molecule towards the ISRF turns out to be around half of the column density seen from the centre of the sphere. We choose shielding factors corresponding to these average column densities. Applying these shielding factors to the unshielded lifetimes derived above, we find that the overall CO lifetime at 85\,au is similar in the clump and the disk: $\sim$50\,a.

\subsubsection{C production timescale}
Using the CO mass $3.4\times10^{-5}$\,M$_{\oplus}$ derived by \citet{Matra_etal_2017}, a CO lifetime of 50\,a leads to a CO mass loss rate of $1.2\times10^{11}$\,kg\,s$^{-1}$. Under the assumptions that 1) the CO mass loss rate is constant, 2) C is produced only from CO photodissociation (C might also be produced from the destruction of other carbon-bearing molecules such as for example methane, but we expect this to be a minor contribution given typical abundances in Solar System comets \citep[e.g.][]{Mumma_Charnley_2011}) and 3) no C is removed from the system, we can use the carbon mass derived in section \ref{sec:simple_mass_estimation} to calculate the time-scale over which CO and C production has been ongoing: $t_\mathrm{CO}=t_\mathrm{C}=N_\mathrm{C}/\dot{N}_\mathrm{CO}$ with $N_\mathrm{C}$ the total number of carbon atoms and $\dot{N}_\mathrm{CO}$ the destruction rate of CO in molecules per second. Using the C mass range derived from our 1D model (section \ref{sec:simple_mass_estimation}), we find a timescale between $1.7\times10^3$ and $1.1\times10^4$ years. When using the mass from the best fitting `ring + clump' model of section \ref{sec:ring_clump}, we find a timescale of $\sim$3\,000 years. These are very short timescales compared to the age of the system. Thus, any event invoked to explain the observed CO clump (e.g.\ a giant collision) needs to occur at a correspondingly high rate---otherwise, it would be unlikely to observe the results of such an event so shortly after it occurred. In the following, we adopt 5\,000\,a as an average estimate of the production timescale.

A caveat remains that some of the C produced from CO photodissociation has already been removed. \citet{Kral_etal_2016} suggested that in steady state, atomic gas is removed by forming an accretion disk inside and a decretion disk outside of the CO-producing parent belt. However, our data argue against such a scenario (see section \ref{sec:no_accretion_disk}). Another possibility is chemical processes that might be able to change the amount of carbon in the disk \citep{Higuchi_etal_2017}. This would require a sufficiently high H$_2$ density. Removing carbon by radiation pressure seems unlikely. First, both neutral and ionised C do not feel any radiation pressure around $\beta$~Pic. However, as shown by \citet{Fernandez_etal_2006}, ions are coupled into a single fluid via Coulomb interactions. But the effective radiation pressure on this fluid is not sufficient for blow out (this explains why e.g.\ Na is seen in Keplerian rotation despite feeling strong radiation pressure), so C is not blown out as part of the fluid either. Recondensation onto dust grains is also irrelevant \citep[][although they considered the recondensation of water, but similar arguments apply for C gas]{Grigorieva_etal_2007}. Finally, accretion by a planet seems unlikely. Gap-opening planets (Jupiter-class) are the best candidates. But even such planets have a finite accretion efficiency (typically 75\%--90\%) limited by the leakage of flow across their gaps \citep{Lubow_DAngelo_2006}. So to increase the estimated lifetime by a factor of $10$ or more, the hypothetical planet would have to sit close to the observed belt and accrete with an efficiency of $90\%$ or higher. In addition, planets down to 1\,M$_\mathrm{Jup}$ have been excluded beyond 30\,au by direct imaging searches \citep{Absil_etal_2013}.

\subsection{Why is there no accretion disk?}\label{sec:discussion_accretion_disk}
\citet{Kral_etal_2016} predicted the formation of a C accretion disk based on thermo-hydrodynamical modelling. However, we showed in section \ref{sec:no_accretion_disk} that no such accretion disk is present. If the C gas production indeed started as recently as estimated in section \ref{sec:C_production_timescale}, the absence of an accretion disk is actually not surprising. Indeed, the timescale for viscous accretion from a radius $r$ can be estimated by 
\begin{equation}
t_\mathrm{vis} = \frac{r^2}{\alpha c_\mathrm{s}H}
\end{equation}
with $c_\mathrm{s}$ the sound speed and $H$ the scale height. We assume $c_\mathrm{s}=\sqrt{\frac{kT}{\mu m_\mathrm{p}}}$ with $T=75$\,K and $\mu=14$, and $H$ as in equation \ref{eq:scale_height} with $r=85$\,au. A very high $\alpha\gtrsim 4$ (taking the upper bound of the timescale range calculated in section \ref{sec:C_production_timescale}) would be required to match the viscous timescale with the time since C production started. In other words, for any reasonable value of $\alpha$, the gas has not yet had enough time to form an accretion disk.

However, a certain amount of carbon is still needed in the inner regions of the disk where metals such as Na or Fe are seen in Keplerian rotation. These species are strongly affected by radiation pressure and C is needed to prevent them from being blown out. As was shown by \citet{Fernandez_etal_2006}, C, which does not feel any radiation pressure around $\beta$~Pic, is acting as a braking agent in the form of C$^+$ via Coulomb interactions. In order to test whether the amount of carbon needed to brake metals is consistent with the new ALMA \ion{C}{1} data, we consider an accretion disk model extending from 10 to 50\,au with a carbon surface density $\Sigma_\mathrm{C}\propto r^{-1}$ and temperature $T\propto r^{-0.5}$ (with $T=75$\,K at 85\,au) \citep{Lynden-Bell_Pringle_1974}. We compute the emission from this model as described in section \ref{sec:3D_modelling}. We then search for \ion{C}{1} emission only in those regions of the datacube where the model predicts emission. However, we want to exclude regions of the datacube that can contain emission from the outer disk. Thus, for those data points with $|v|<v_\mathrm{orb}(50\mathrm{au})$ (with $v_\mathrm{orb}(r)$ the orbital speed at radius $r$), we request points to be at least one spatial resolution element inside of the line in the pv diagram defining a thin ring at 50\,au (see figure \ref{fig:pv_diagram}). For points with a larger $|v|$, we can be sure that no contamination from the outer disk is present. We also exclude points with a predicted emission below 10\% of the model peak to avoid considering regions of the datacube where only weak emission is expected anyway. We can then measure the flux in this region of the datacube, with the error estimated with a method analogous to what is described in appendix \ref{sec:error_calculation_details}. Defining a detection threshold of $3\sigma$, we do not detect significant emission. We derive an upper limit on the C density by adjusting the model such that the probability for a non-detection, i.e.\ measuring a flux below the detection threshold, would only be 1\% if the model was correct. Assuming Gaussian noise, the probability of a model with flux $F_\mathrm{mod}$ to remain undetected is given by $\frac{1}{2}\left[1+\erf\left( \frac{n\sigma-F_\mathrm{mod}}{\sqrt{3}\sigma}\right) \right]$ with $\sigma$ the error on the flux and $n=3$ in our case. The upper limit model has a mid-plane C$^+$ number density of $\sim$600\,cm$^{-3}$ at 30\,au, while $\sim$100\,cm$^{-3}$ are necessary to brake the metals \citep{Fernandez_etal_2006}. Thus, the data are consistent with enough carbon being present in the inner disk to brake metals. However, to get better constrains, we also look at the \ion{C}{2} observations from \textit{Herschel}/HIFI by \citet{Cataldi_etal_2014}. We measure the flux in the wings of the \ion{C}{2} spectrum corresponding to radial velocities between $v_\mathrm{orb}(50\mathrm{au})$ and $v_\mathrm{orb}(10\mathrm{au})$. Errorbars are calculated by calculating the flux in spectral regions of the same size without line emission and taking the standard deviation. We first detemine the errors on the flux in the left wing and right wing individually and then add them in quadrature to obtain the error on the total measured flux. No significant (larger than 3$\sigma$) flux is detected neither in the H nor V beam of the data. Thus, we adjust the model such that the combined probability to remain undetected in the ALMA \ion{C}{1} and the HIFI \ion{C}{2} (H and V beam) data is only 1\% (including the ALMA data has a negligible effect as the HIFI data are more constraining). This model has a mid-plane C$^+$ density at 30\,au of $\sim$380\,cm$^{-3}$. We conclude that the currently available data are consistent with carbon being present inside of 50\,au at a level that is sufficient to brake metals. We suggest that this gas in the inner region was not produced in the same event as the gas seen at larger distances. If it was, we would expect a full accretion disk, which is not supported by the data \citep[see also][]{Cataldi_etal_2014}. Instead, it might be the leftover from a previous gas-producing event.
%here I should perhaps also discuss what the absence of the accretion disk implies with respect to the Xie et al. (2013) study.

\subsection{Origin of the observed C asymmetry}\label{sec:C_asymmetry_origin}
A key result from our new ALMA data is that C shows the same asymmetry as CO: a clump on the SW side of the disk. This is surprising since one would expect C to spread in azimuth within a few orbits even though C production might happen preferably at the CO clump. This is in contrast to CO which has a lifetime shorter than an orbital period and thus remains asymmetric. Here we discuss possible solutions to this puzzle.

\subsubsection{Recent event}
\begin{figure}
\plotone{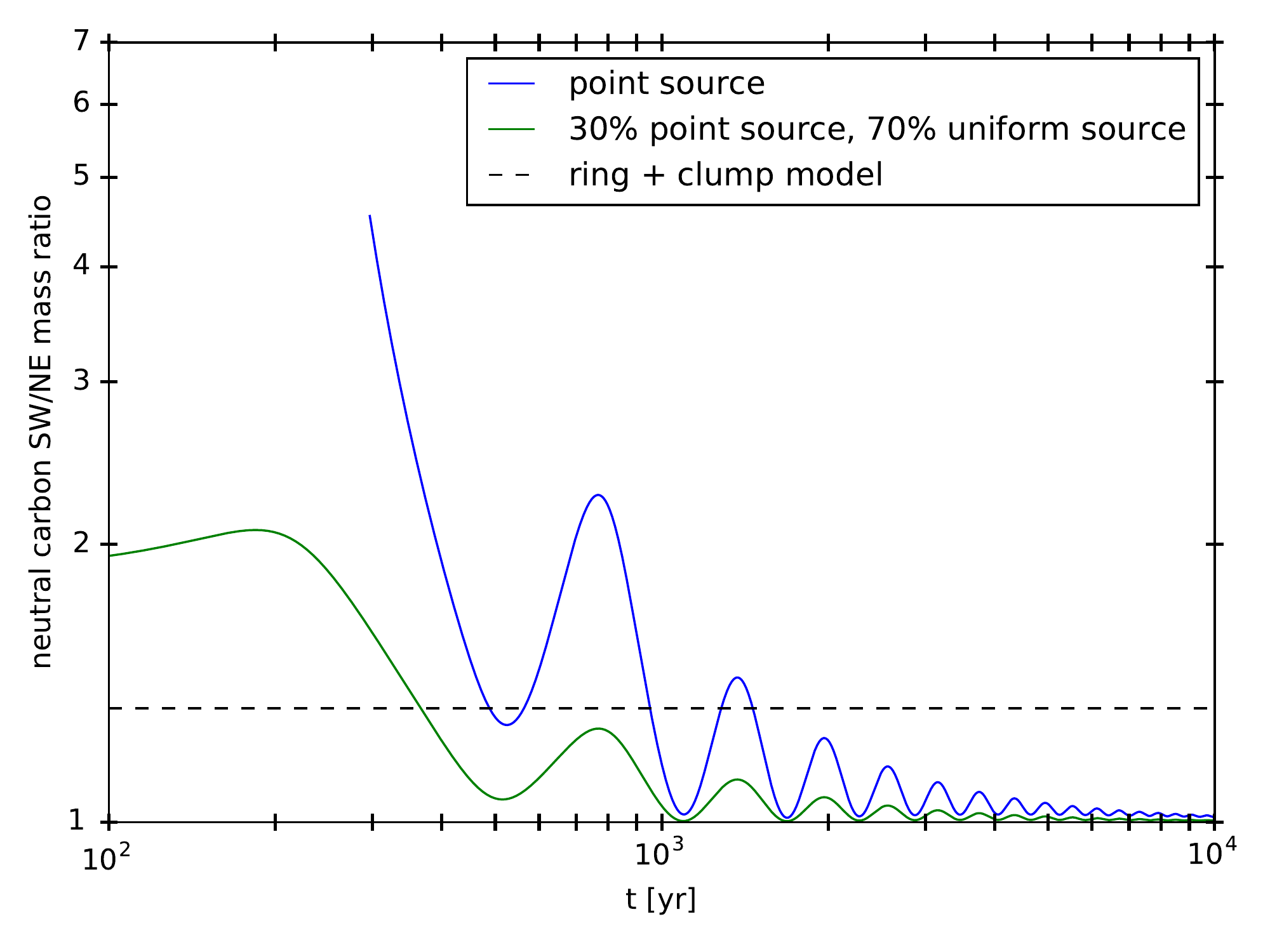}
\caption{Temporal evolution of the NE/SW ratio of the neutral carbon mass in our toy model. Two scenarios are considered: production from a point source (blue line), providing an upper limit on the symmetrisation timescale, and a scenario with initial conditions similar to what we observe today (green line). The dashed line denotes the mass ratio of our best-fit `ring+clump' model.\label{fig:ne2sw_ratio_evolution}}
\end{figure}

Perhaps the simplest explanation for the observed C asymmetry is that C production at the clump location started so recently that there was not yet enough time for azimuthal spreading (respectively symmetrisation). We expect the timescale for azimuthal symmetrisation to be on the order of a few orbits (at 85\,au, the orbital period is $\sim$600 years). We investigate the symmetrisation with a 1D toy model, where the only dimension is the azimuthal angle $\phi$. To start, we assume that all C is produced in a single point at azimuth $\phi_0$, at a distance of 85\,au, with a rate equal to the CO destruction rate calculated in section \ref{sec:C_production_timescale}. In reality, only 30\% of the CO emission is found in the clump \citep{Dent_etal_2014}. This setup thus maximises the asymmetry, so the derived symmetrisation timescale can be considered an upper limit. We then write the following simple equations describing the temporal evolution of the neutral and ionised carbon densities under the influence of C production, ionisation, recombination and orbital motion:
\begin{eqnarray}
\frac{\partial \rho_0}{\partial t} &=& \delta(\phi-\phi_0)\Lambda_\mathrm{CO}\frac{m_\mathrm{C}}{m_\mathrm{CO}} + \rho_+n_\mathrm{e}\gamma - \rho_0\Gamma - \omega\frac{\partial\rho_0}{\partial\phi}\label{eq:C0_azimuth}\label{eq:toy_model_neutral}\\
\frac{\partial \rho_+}{\partial t} &=& -\rho_+n_\mathrm{e}\gamma + \rho_0\Gamma - \omega\frac{\partial\rho_+}{\partial\phi}\label{eq:toy_model_ionised}
\end{eqnarray}
where $\rho_0$ and $\rho_+$ are the azimuthal densities (in kg\,rad$^{-1}$) of neutral and ionised C respectively, $\Lambda_\mathrm{CO}$ is the CO destruction rate (in kg\,s$^{-1}$), $m_\mathrm{C}$ and $m_\mathrm{CO}$ are atomic and molecular masses respectively, $n_\mathrm{e}$ the electron density, $\gamma$ the recombination coefficient, $\Gamma$ the ionisation rate and $\omega$ the angular velocity at 85\,au. For the electron density, we assume that all electrons come from the ionisation of C, and that the typical volume extends over $\Delta r=$70\,au in the radial direction (best-fit `ring' model, section \ref{sec:ring}) and over $\Delta z=2H$ (with $H$ the scale height, see equation \ref{eq:scale_height}) in the vertical direction (the model remains one-dimensional; we only assume a certain volume to get a value for the electron density). This means that $n_\mathrm{e}=\rho_+/(m_\mathrm{C^+}r\Delta r\Delta z)$ where $m_\mathrm{C^+}$ is the mass of a C$^+$ ion. Then, equations \ref{eq:toy_model_neutral} and \ref{eq:toy_model_ionised} are used to compute the change of the azimuthal densities at each time step. Figure \ref{fig:ne2sw_ratio_evolution} shows how the SW/NE mass ratio of neutral carbon evolves with time. The wavy pattern is due to the orbital motion of the gas. The local minima correspond to the times when the first gas produced by the point source leaves the NE side and enters the SW side. After $\sim$10$^3$ years, the ratio falls below the value of our best-fit `ring + clump' model.

We can also use this model to estimate the time it will take to symmetrise the C from the state that what we observe today if there is no mechanism that prevents symmetrisation. For example, assuming that only 30\% of the C production occurs at the clump, with the other 70\% uniformly distributed at all azimuths (this corresponds to adding a production term independent of $\phi$ to equation \ref{eq:C0_azimuth}), we get the green curve shown in figure \ref{fig:ne2sw_ratio_evolution}. As expected, the symmetrisation occurs faster---within $\sim$10$^3$ years, the SW/NE mass ratio falls below 1.1.

A caveat to this toy model is the possibility that the gas production is not constant over time. Also, for more robust constraints, detailed hydrodynamical calculations would be necessary.

The upper limit on the symmetrisation timescale is below the lower bound of the timescale range over which C production occurred, as derived from the C mass (section \ref{sec:C_production_timescale}). Thus, it appears unlikely that the asymmetry can be explained by invoking a very recent event.

In summary, the estimated lifetime ($\sim$5\,000\,a) of the carbon-rich gas disk should be long enough to spread out the azimuthal asymmetry, but not long enough to diffuse the disk radially via viscous spreading.

\subsubsection{Resonance trapping by planet}

\citet{Matra_etal_2017} argued that the asymmetry in $\beta$~Pic, and perhaps in a number of other debris disks, is the result of planetesimals being trapped into mean-motion resonances (MMR) by a (migrating) planet. For $\beta$~Pic, our resolved \ion{C}{1} observation excludes this possibility.

For planetesimals in resonance with a planet there would be an enhancement in particle density at some special azimuth relative to the planet. In the case of 2:1 MMR, such a resonance island lies $90\arcdeg$ behind and ahead the azimuth of the perturbing planet. As collisions are more frequent in the denser region, it is expected, in this scenario, that CO, which essentially traces recent collisions because of its short lifetime, is enhanced near the island and therefore asymmetric. However, as the planet revolves in its orbit, the resonance island sweeps through all azimuths. Integrated over many periods, the resonance island does not linger particularly long around a special azimuth. So if we look at a tracer gas that is the integrated production of CO over many periods, as the carbon gas is (produced over $\sim$5\,000 \,a, i.e.\ $\sim$10 orbits), we expect it to be evenly spread out in the orbit. The observed asymmetry in C thus cannot be explained by its parent planetesimals being trapped into a planet MMR.

Can carbon gas also be trapped into a MMR? In contrast to dust grains, gas cannot be trapped into an MMR. One can show that the resonance width, even when forced by a highly eccentric Jovian planet, is narrower than the vertical scale height of the gas ($\sim$ a few au). As the vertical scale height corresponds to the sound speed travel time over an orbit, gas pressure can, in an orbital period, easily disperse the gas azimuthally and radially over an extent wider than the resonance width. It is difficult for the weak planetary perturbation to restrain them.
% width = 2 au \sqrt(mu/1.0e-3 * e/0.5) (a/85au)
% vertical scale height ~ 0.05a ~ 4 au

In section \ref{sec:eccentric_gas_distribution} we showed that an eccentric gas distribution is qualitatively able to reproduce the carbon observations. Thus, in the following sections, we discuss how such an eccentric disk could arise.

\subsubsection{Initially eccentric disk}
If the parent planetesimal disk is eccentric because it has been left in that state by some unknown initial condition, the debris disk will initially be asymmetric. Such a disk, if precessed rigidly, can retain its initial configuration over a time much longer than the sound crossing time. However, over the lifetime of the system (20\,Ma), differential precession should have disturbed this initial imprint. Ignoring the presence of other planets, and only considering the quadrupole precessional effect of $\beta$\,Pic\,b \citep{Lagrange_etal_2009}, the timescale for order unity change in the relative precession angle is \citep{Murray_Dermott_2000}
% take M_p = 13 M_Jup, take a_p = 8 au, take a_disk=85 au, take delta a=20 au
\begin{equation}
t_{\rm smear} \sim \frac{M_*}{M_{\mathrm{p}}} \left(\frac{a}{a_{\mathrm{p}}}\right)^2
	\left(\frac{2\Delta a}{a}\right)^{-1} P_{\rm orb} \sim 14\,\mathrm{Ma},
\end{equation}
where we evaluate the expression using $M_{\mathrm{p}} = 13$\,M$_{\mathrm{Jup}}$ \citep{Morzinski_etal_2015} for the planet mass and $a_{\mathrm{p}} = 9$\,au \citep{Wang_etal_2016} for its semi-major axis with the ring at $a \sim 85$\,au and with a fiducial radial width of $\Delta a \sim 20$\,au (consistent with that in Table \ref{tab:parameter_space_eccentric_orbits}). So even without an additional planet, an initially eccentric disk is expected to be largely smeared out.

\subsubsection{Eccentric disk secularly forced by a planet}
Secular perturbation from a planet can not be responsible for producing the eccentric gas disk that we observe. The age of the C gas ($\sim$5\,000\,a) is shorter than any reasonable secular timescale ($\geq P_{\rm orb}/\mu$, where $\mu$ is the mass ratio of planet to star).  So any eccentricity in the gas disk will have to be inherited from their planetesimal parent bodies. But if so, what could possibly have started the parent bodies' grind-down a mere $5\,000$\,a ago, if they have lived in such states for a secular timescale? But let us ignore this issue here and proceed to consider the geometry.
%\w{Moreover, to make the secular timescale shorter than the 20 Myrs %lifetime, the planet would have to be more  massive than $\sim 0.1 %M_J$.}
% if a_p = a_dust, t_sec=600 yrs/mu, mu(0.1 M_J) = 6e-5

In the hypothetical case of long-lived parent bodies forced to a relatively low eccentricity, 
% differential precession not an issue. 
more particles will be found near apoaps where they move the slowest \citep[the so-called `apo-centre' glow,][]{Pan_etal_2016}, while collisions are more likely to occur near the pericentre where the particle streamlines are more densely spaced and their relative velocities are higher. Collisions near the pericentre occur at a higher relative velocity, allowing smaller debris (which are more populous and have a larger collisional surface area) to break apart a given fragment. \citet{Matra_etal_2017_Fomalhaut} suggest that the mass loss rate is enhanced at either periaps or apoaps depending on the proper eccentricity and strength of the planetesimals. From preliminary numerical simulations, we favour an enhancement at periaps (more detailed simulations will be presented in a forthcoming paper). As a result, we expect CO (which reflects the instantaneous collisional mass loss rate) to be concentrated in periaps (the SW side, as is observed), while both the submm emission and the carbon line fluxes should be determined mostly by particle trajectories and should peak at apoaps, contrary to the observations (the observed \ion{C}{1} periapsis to apoapsis flux ratio is $1.2\pm0.2$, see section \ref{sec:results_line_emission}). This argument is easy to understand if one thinks of the CI gas as exact analogue of the small dust grains, which are also integrated products of previous collisions. Small grains will shine more brightly in apo-centre because there are more of them there---unless their eccentricities are so large that the fall-off of stellar-flux at apo-centre is more important than the number excess, see our discussion below.

At higher eccentricities, the apocentre is much further than the pericentre, and the drop-off in stellar flux, particle volume density and gas temperature could compensate for the above trajectory effect and reduce the apocentre glow. Colder submm particles have reduced emissivity. The lower gas temperature and volume density also mean that the carbon gas is less excited, leading to reduced line fluxes. This effect is more severe when the orbits have an intrinsic spread in eccentricity, leading to a spread in apocentre distances and further reducing the dust and gas volume densities in those locations. Carbon ionization fraction, on the other hand, may also be affected by the lower ionizing flux and the smaller recombination rate.

So to explain the observed same-sided asymmetry in different tracers, we will require a medium to high eccentricity ($e \gtrsim 0.3$, table \ref{tab:parameter_space_eccentric_orbits}) and preferably a large spread in eccentricities (discussed below). For a secularly forced ring, the forced eccentricity $e \sim 5/4 (a_{\mathrm{p}}/a_{\mathrm{ring}}) e_{\mathrm{p}} $, where $e_{\mathrm{p}}$ is the eccentricity of the planet \citep{Murray_Dermott_2000}. A planet with considerable eccentricity is then required, which may itself decimate the planetesimal belt by dynamical ejection.
% stability + 5/4 alpha e constraint
%{\w I could present a model, however, it takes quite a bit to explain
%  the model set-up, e.g., how luminosity is calculated, how collision
%  rate is calculated...}

Another argument against the planet hypothesis comes from the range of eccentricities required to explain the observations. For a disk with a $20\%$ spread in semi-major axis ($\Delta a \sim 20$\,au if $a =85$ au), we expect a $20\%$ spread in the forced eccentricities. This is too small to help explain the same-side asymmetry and it is also smaller than indicated by our best-fit solution.

In conclusion, it seems the hypothesis that an underlying highly-eccentric planetesimal belt, forced secularly by a planet, is unlikely to explain our observations.

\subsubsection{Giant Impact}
\citet{Jackson_etal_2014} proposed a scenario where a giant impact between two comparably massed bodies produces a wide spread of debris that have a range of eccentricities but a similar alignment. This is qualitatively plausible to explain the observations \citep[but see][for a counter argument regarding the radial width]{Matra_etal_2017}. If such an event occurred in the recent past, it provides an interesting explanation for our deduced carbon production timescale.
However, such an event seems exceedingly improbable, as we will show with an order-of-magnitude estimates of the event rate. 

Consider $N$ bodies with radius $R$, in a belt of semi-major axis $a$ and width $\Delta a$,  performing vertical epi-cycles around the mid-plane. Viewed by each one of these bodies, the other bodies present a certain optical depth of $\tau_\perp = (N 4 \pi R^2)/(2\pi a \Delta a)$, or a mean collision time of $P_{\rm orb}/\tau_{\perp}$.  Summed over $N$ bodies, this implies a mean event time of
\begin{equation}
t_{\rm coll} \sim P_{\rm orb} \frac{2 \pi a \Delta a}{N^2 \times 4 \pi
    R^2} \sim 3\times 10^{9} {\rm \, yrs} \times \left({\frac{a}{80 {\rm\,au}}}\right) 
    \left(\frac{\Delta a}{20 {\rm\,au}}\right) \left(\frac{N}{1\,000}\right)^{-2}
\left(\frac{R}{2\,000{\rm\,km}}\right)^{-2}\, ,
\label{eq:chance}
\end{equation}
where we assume there is no gravitational focusing among these low-mass objects, reasonable if their velocity dispersion lies much beyond their surface escape velocities. The scaled value $R=2000$km is appropriate to explain the total gas/dust mass observed, and $\Delta a = 20$ AU is inspired by the best fit in Table \ref{tab:parameter_space_eccentric_orbits}. The value $N=1000$ is a place-holder. So, to produce an event as recent as 5\,000\,a ago, one would require $N \sim 10^5$, or an absurdly high total mass of $\sim 500 M_\oplus$. This argument makes giant impacts exceedingly implausible at this location in the disk.  Another difficulty with such a scenario is that giant impacts tend to be completely accretionary at low relative speeds, and even at speeds beyond the surface escape, only a small fraction of mass can be unbound and released into the circumstellar environment \citep{Agnor_Asphaug2004}.

\subsubsection{Tidal Disruption}\label{sec:tidal_disruption}

Here, we briefly propose an alternative scenario to produce the observed disk. A more detailed calculation will be presented in an upcoming paper.
% 12 r_Hill ~ 30au
Consider that the outer disk contains a number of Neptune-like planets ($N_N$ $\sim$ a few). They have a surface escape velocity of $v_{\mathrm{esc}} \sim 23$\,km\,s$^{-1}$. Let us also assume that there are $N$ bodies similar in mass to the Moon ($\sim 0.01 M_\oplus$) or Mars ($\sim 0.1 M_\oplus$) and moving in space with a relatively low dispersion velocity, $\sigma \ll v_{\mathrm{esc}}$. There is little gravitational focussing among these bodies, but strong focussing by the Neptunes. The timescale for a physical collision with a Neptune is
\begin{equation}
t_{\rm coll} \sim P_{\rm orb} 
 \frac{2 \pi a \Delta a}{N N_N \times \pi
    R_N^2} \left(\frac{\sigma}{v_{\rm esc}}\right)^2
\sim 3\times 10^{7}\, 
{\rm\,yrs} \times\left({{\frac{a}{80{\rm\,au}}}}\right)
\left({\frac{\Delta a}{20{\rm\,au}}}\right)
\left(\frac{N}{1\,000}\right)^{-1} \left(\frac{\sigma}{1\,\mathrm{km\,s}^{-1}}\right)^{2}
	\left(\frac{v_{\rm esc}}{23\,\mathrm{km\,s}^{-1}}\right)^{-2}\, .
\label{eq:chance2}
\end{equation}
where we chose $N_\mathrm{N}=5$. This is somewhat arbitrary, but the abundance of cold Neptunes is already suggested by micro-lensing studies, which claim that their abundance rate per star (mostly M-dwarfs) is 52\% \citep{Cassan_etal_2012}.
This timescale is becoming astrophysically interesting. But a physical collision with a Neptune will not produce a debris disk around the star. Instead, we focus on encounters that are close enough that the body is tidally disrupted. This means encounter distance of $\sim 2 R_N$ and the encounter time $t_{\rm disruption} \sim t_{\rm coll}/2 \sim 10$\,Ma (at $2 R_N$, while the geometric cross-section goes up by a factor of $4$, the gravitational focussing factor, $(\sigma/v_{\rm esc})^2$, evaluated at $R = 2 R_N$ goes down by a factor of $2$).
% the linear scaling comes about because focussing is also weaker by
% one radius factor
So there could have been a few tidal disruption events over the lifetime of $\beta$~Pic, if there are thousands of moons floating around and if these moons retain low enough dispersion velocities to experience strong gravitational focussing by Neptune. This is still well above the 5\,000\,a event time we infer and still presents a tension.

The end product of the tidal disruption shares many similarities with that from a giant impact. First, the debris will be ejected on a variety of orbits, bringing about a large spread in eccentricity. The semi-major axis will also have a spread.  All debris will return to the disruption site, making this location a region of frequent collisions. The size of the disruption site, however, is larger than that in giant impact. As the debris flies away from Neptune, its stellar-centric orbit is altered by Neptune's gravity while it is still within Neptune's Hill sphere. As a result, unlike the narrow nozzle of the size of the impactor radius in the case of giant impacts, here, the nozzle has a typical spread of order the planet's Hill radius, which reduces the peak collision rate. However, \citet{Matra_etal_2017} measured a radial extent of the CO clump of at least 100\,au, which is clearly larger than the planet's Hill sphere. More detailed modelling is needed to see whether a tidal disruption event can produce such a radially broad clump. Also, the clump in the deprojection of \citet{Matra_etal_2017} might appear more extended than it is in reality because the intrinsic velocity dispersion of the gas and the finite velocity resolution of the instrument degrade the resolution of the deprojection in the $y$ direction. In addition, the deprojection is carried out assuming velocities corresponding to circular orbits. Thus, if the orbits are eccentric in reality, the deprojection will be distorted and not correspond to the actual gas distribution.

The event in $\beta$~Pic is recent and must also be relatively short-lived, assuming we are not observing it at a special time. The duration of a tidal disruption flare will depend on the above-discussed collisional geometry, but also on the distribution of the largest fragments (that remain or reform) after the disruption event. Both of these effects need to be studied in detail. In addition, a short lifetime will also ensure that debris products from previous events do not interfere with the current event. If not, the asymmetry would be washed out by previous debris which likely have a different asymmetry; and we would observe the radially diffused accretion disk from gas produced in previous events.

In summary, our observation disfavours a few proposed scenarios (planet MMR trapping, giant impact, secular forcing). Instead, we propose that the bright, asymmetric debris disk in $\beta$~Pic could be the result of a recent tidal disruption of a Moon to Mars-sized object by a Neptune-like planet. Using the C mass derived in section \ref{sec:simple_mass_estimation}, and assuming that the disrupted body has had a CO mass fraction of 10\%, its total mass would indeed be $\gtrsim$3\,$M_\mathrm{Moon}$ (lower limit because not all carbon might have been released yet).

\subsection{Comparison with the CO emission}\label{sec:CO_modelling}
\begin{figure}
\plotone{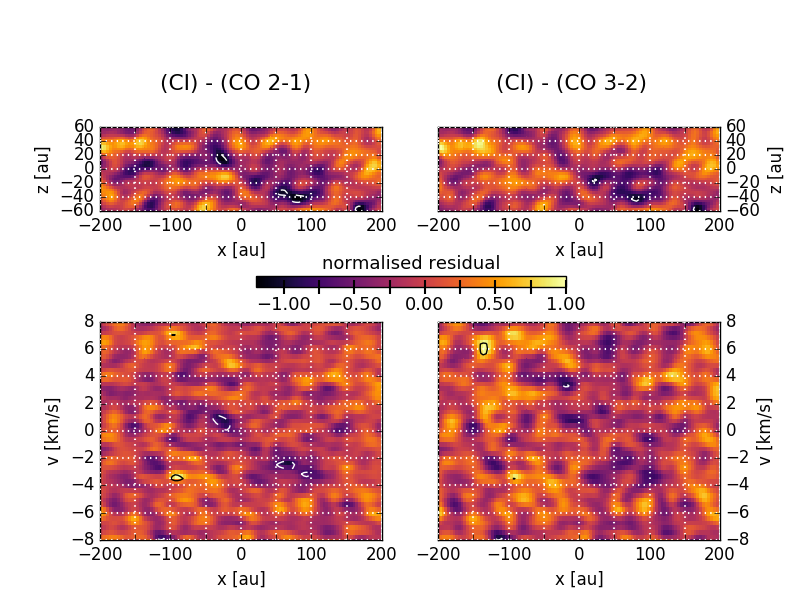}
\caption{Residual moment 0  and pv diagram when subtracting the peak-scaled CO from the \ion{C}{1} data cube. The spectro-spatial resolution has been adjust prior to subtraction. Contours are drawn at intervals of 3$\sigma$. The residual data cube was integrated over $\pm$6\,km\,s$^{-1}$ for the moment 0 and $\pm30$\,au for the pv diagram respectively.\label{fig:CO_emission_residuals}}
\end{figure}

Since C and CO have similar horizontal emission profiles (see Figure \ref{fig:mom0_xprofile_pv_restriction}), the question arises how similar the spatial and spectral distribution of the emission really is. To answer this question, we first interpolate the CO data cube onto the coordinates of the \ion{C}{1} data cube. Then, we use convolution to adjust the spatio-spectral resolution of the data cubes. Finally, we normalise the data cubes and subtract. Figure \ref{fig:CO_emission_residuals} shows the moment 0 and pv diagram of the residual cube for the CO~2--1 and 3--2 transitions \citep{Matra_etal_2017}. The CO emission seems similar to the \ion{C}{1} emission, with few residuals above 3$\sigma$ seen. With the data at hand, we are thus not able to detect clear differences in the emission distribution. Note that there is significant difference in the distribution of CO~2--1 vs CO~3--2 emission \citep{Matra_etal_2017}.

\subsection{Comparison to the spatial distribution of metals}\label{sec:metals_spatial_distribution}
\citet{Brandeker_etal_2004} and \citet{Nilsson_etal_2012} found that Na and Fe have a NE/SW asymmetry reminiscent of what we see for C and CO \citep[compare figure \ref{fig:mom0_xprofile_pv_restriction} of this work to figure 3 in][]{Brandeker_etal_2004}. Na and Fe also show a tilt similar to what is seen for C and CO. The radial distribution is quite different, however, with the density being higher closer to the star instead of peaking at 85\,au.  The asymmetry is seen in both the inner and outer parts of the disk; concerning the inner distribution of Na and Fe, the emission is traced inwards to the observational limit of 13\,au in the NE, while to the SW the density peak appears to be located much further out, beyond 50\,au. In the outer parts of the disk, Na and Fe can be seen all the way to the limit of the observation at the projected distance of 330\,au in the NE, while the disk seems truncated in the SW at 150--200\,au.

A possible scenario is that the origin of CO (and thus C and O as its dissociation products) is different from the origin of the metals observed in the optical \citep[Na, Fe, Ca, Ni, Ti, and Cr,][]{Brandeker_etal_2004}. While CO likely comes from the disruption of CO-rich cometary bodies at 85\,au, the metals observed in the optical could be produced by the so-called falling evaporating bodies \citep[FEBs, e.g.][]{Vidal-Madjar_etal_2017} close to the star and then diffused outwards by the stellar radiation pressure. As the presence of C$^+$ would act as a braking agent \citep{Fernandez_etal_2006}, its spatial distribution would naturally become imprinted on the spatial distribution of the metals. The spatial distribution of Na and Fe is therefore consistent with C$^+$ being in an eccentric distribution resulting from a tidal disruption event, as outlined in sections \ref{sec:eccentric_gas_distribution} and \ref{sec:tidal_disruption}. With the SW clump at 85\,au being at the convergence point for a family of eccentric orbits, the C would be more concentrated to the clump location in the SW, while being much more radially distributed in the NE, in agreement with the observed Na and Fe distributions.

\section{Summary and conclusions}\label{sec:conclusion}
In this paper, we present resolved ALMA band 8 observations of \ion{C}{1} emission towards the $\beta$~Pic debris disk. Our work can be summarised as follows:
\begin{enumerate}
	\item Using a simple 1D model that calculates the ionisation balance and non-LTE level populations, we estimate the total C mass to be between $5\times10^{-4}$ and $3.1\times10^{-3}$\,M$_\earth$. Assuming that C is produced from the photodissociation of CO at a constant rate, and that C is not removed from the system, this mass implies that gas production started only $\sim$5\,000\,a ago.
	\item Surprisingly, \ion{C}{1} shows the same asymmetry as seen for CO: a clump on the SW side of the disk. By modelling the spatial distribution of the C gas, we find that a satisfactory fit to the \ion{C}{1} and archival \ion{C}{2} data can be found by assuming that the gas consists of a ring between 50 and 120\,au with a superimposed clump at the same location as the CO clump. A model assuming eccentric orbits of the gas with a flat eccentricity distribution between 0 and 0.3 also reasonably fits the data.
	\item The \ion{C}{1} data are not consistent with the accretion disk predicted by \citet{Kral_etal_2016}. If C gas production indeed started only $\sim$5\,000\,a ago, not enough time has passed for gas to have spread viscously into an accretion disk.
	\item The short timescale since gas production started argues against a giant impact-origin of the C/CO/dust clump, because giant impacts are rare. It is unlikely that we observe the results of such an event so shortly after it occurred. However, while the production timescale of $\sim$5\,000\,a is short compared to the age of the system, it is long enough to allow azimuthal spreading of the gas. Thus, a scenario where the C assymetry is due to a lack of time for azimuthal spreading is disfavoured.
	\item The fact that C shows the same asymmetry as CO (a clump in the SW) argues against a scenario where the clump is due to planetesimals trapped in a resonance with an outward-migrating planet. Indeed, in such a scenario, the clump orbits with the planet, i.e.\ the gas production should be symmetric on an orbital timescale.
	\item In order to explain the simultaneous C and CO asymmetry, we propose that the planetesimal and gas disk of $\beta$~Pic is eccentric and might have originated from a recent tidal disruption event. This could potentially also explain the asymmetry observed in \ion{Na}{1} and \ion{Fe}{1} by \citet{Brandeker_etal_2004}.
\end{enumerate}
A detailed study of the tidal disruption mechanism will be presented in a forthcoming paper. Follow-up observations of the \ion{C}{1} line at higher SNR and spectro-spatial resolution can be used to confirm or reject our hypothesis of an eccentric disk due to a tidal disruption event.

%
%The first step to making this happen is to have the data or software in
%a long term repository that has made these items available via a persistent
%identifier like a Digital Object Identifier (DOI).  A list of repositories
%that satisfy this criteria plus each one's pros and cons are given at \break
%\url{https://github.com/AASJournals/Tutorials/tree/master/Repositories}.
%
%In the bibliography the format for data or code follows this format: \\
%
%\noindent author year, title, version, publisher, prefix:identifier\\
%
%\citet{2015ApJ...805...23C} provides a example of how the citation in the
%article references the external code at
%\url{http://dx.doi.org/10.5281/zenodo.15991}.  Unfortunately, bibtex does
%not have specific bibtex entries for these types of references so the
%``@misc'' type should be used.  The Repository tutorial explains how to
%code the ``@misc'' type correctly.  The most recent aasjournal.bst file,
%available with \aastex\ v6, will output bibtex ``@misc'' type properly.

\acknowledgments
We are grateful to the anonymous referee for his/her exceptionally thorough review, leading to a significant improvement of our manuscript. We would like to thank S\'ebastien Muller and Ivan Mart\'i-Vidal from the Nordic ARC node for their crucial support in calibrating the data. We acknowledge helpful discussions with Thayne Currie, Quentin Kral, Luca Matr\`a, Nagayoshi Ohashi, Alfred Vidal-Madjar and Mark Wyatt. Luca Matr\`a kindly provided the ALMA CO 3--2 and 2--1 data as well as the model of the dust radiation field. Quentin Kral helped getting the \texttt{LIME} code running. This work was partially supported by JSPS KAKENHI Grant Number JP16F16770. This work was supported by the Momentum grant of the MTA CSFK Lend\"ulet Disk Research Group. This paper makes use of the following ALMA data: ADS/JAO.ALMA\#2013.1.00459.S. ALMA is a partnership of ESO (representing its member states), NSF (USA) and NINS (Japan), together with NRC (Canada), MOST and ASIAA (Taiwan), and KASI (Republic of Korea), in cooperation with the Republic of Chile. The Joint ALMA Observatory is operated by ESO, AUI/NRAO and NAOJ. This research has made use of NASA's Astrophysics Data System.


\vspace{5mm}
\facilities{ALMA, Herschel (PACS, HIFI)}


\software{astropy \citep{Astropy_etal_2013}, CASA \citep{McMullin_etal_2007}, emcee \citep{Foreman-Mackey_etal_2013}, LIME \citep{Brinch_Hogerheijde_2010}, Matplotlib \citep{Hunter_2007}, NumPy \citep{vanderWalt_etal_2011}, pythonradex (\url{https://github.com/gica3618/pythonradex}), SciPy (\url{http://www.scipy.org})}

\appendix
\section{Details of the error calculation}\label{sec:error_calculation_details}
In this section we describe how the errors quoted in section \ref{sec:results_line_emission} and \ref{sec:continuum} are derived.

The total line emission was measured by integrating the data cube within $\pm$6\,km\,s$^{-1}$ in the spectral dimesion and over the rectangular box shown in Figure \ref{fig:mom0_xprofile} in the spatial dimension. Having measured the noise in the data cube, a naive way to estimate the error $\sigma_\mathrm{F}$ on the total flux would be to write $\sigma_\mathrm{F}=\sqrt{n}\sigma\mathrm{d}\Omega\mathrm{d}v$ with $n$ the number of data points over which the integration extends, $\sigma$ the noise in the data cube and $\mathrm{d}\Omega$ and $\mathrm{d}v$ the extent in solid angle and velocity of a single data point. However, since neighboring data points are correlated, this approach is not valid. Instead, we consider the flux of a number of volumes with the same size as the volume used to measure the total flux, placed in regions of the data cube without emission. Taking the standard deviation of these flux samples, we obtain an estimate of the error. The volumes are placed such that they are sufficiently distant from each other to be considered independent. Since no primary beam correction has been applied, the noise can be assumed uniform over the data cube.

For the continuum, the above procedure yields too few flux samples. Thus, we reduce the size of the flux samples in the $x$ direction to get more samples. Then, we set $\sigma_\mathrm{F}=\sigma_\mathrm{s}\sqrt{N}$ where $\sigma_\mathrm{s}$ is the standard deviation of the flux samples and $N$ is the number of flux samples that fit into the region for which the flux is measured. Again, the flux samples are placed sufficiently distant from each other to be considered independent.

For the emission profile along the disk (Figures \ref{fig:mom0_xprofile} and \ref{fig:mom0_xprofile_pv_restriction} right), we consider sample `volumes' that extend only one pixel in the $x$ direction. In the case of the profile derived from the restricted region in pv space (Figure \ref{fig:mom0_xprofile_pv_restriction}), the size of the region over which we integrate depends on $x$. Thus, we take flux samples for each $x$ individually. For this profile, the error thus depends on $x$.

\section{Measurement of the line broadening parameter}\label{sec:line_broadening}
The line broadening parameter $b$ (defined as $b=\mathrm{FWHM}/(2\sqrt{\ln 2})$ where $\mathrm{FWHM}$ is the full width at half maximum of the line) parametrises the line broadening due to effects other than the orbital motion of the gas (e.g.\ thermal broadening or turbulence). We can use the ALMA observations of C and CO to measure $b$ by considering a vertical cut in the PV diagram (Figure \ref{fig:pv_diagram}) going through $x=0$, i.e.\ considering the line of sight towards the star. For circular orbits, all gas along this line of sight is centred at the same radial velocity (namely 0\,km\,s$^{-1}$), allowing a measurement of $b$. However, if orbits are eccentric, the line of sight towards the star can contain additional broadening due to orbital motion. Furthermore, the finite spatial resolution of the instrument will blur material with non-zero radial velocity into the line of sight towards the star. Thus, our derived value of $b$ should be considered an upper limit.

We consider the PV diagrams of \ion{C}{1} (this work), CO~2--1 and CO~3--2 \citep{Matra_etal_2017} and for each of them fit a Gaussian to the vertical cut through $x=0$, yielding three independent measurements of $b=\sqrt{b_\mathrm{fit}^2-b_\mathrm{inst}^2}$ where $b_\mathrm{fit}$ is the broadening parameter of the fitted Gaussian and $b_\mathrm{inst}$ is the broadening due to the spectral response function of the instrument\footnote{See ALMA Cycle 6 Technical Handbook, section 5.5.2 (Spectral Setup), \url{https://almascience.nao.ac.jp/documents-and-tools/cycle6/alma-technical-handbook}}. We estimate errors using \texttt{emcee}, a \texttt{python} implementation of an affine invariant Markov chain Monte Carlo (MCMC) sampler \citep{Foreman-Mackey_etal_2013}. The likelihood was defined via a $\chi^2$ measure analogous to equations 1 and 2 by \citet{Booth_etal_2017}. In particular, we also use a noise correlation ratio, defined in our case as the square root of the number of spectral pixels per spectral resolution element. We consider invariant, uninformative priors, imposing only that the peak of the Gaussian and the broadening parameter are positive. We use 100 walkers with $10^4$ steps. This yields the following results for $b$: $0.66\pm0.11$\,km\,s$^{-1}$ (\ion{C}{1}), $0.56\pm0.04$\,km\,s$^{-1}$ (CO~2--1) and $0.56\pm0.05$\,km\,s$^{-1}$ (CO~3--2). Combining these measurements as a weighted mean yields $b=0.57\pm0.03$\,km\,s$^{-1}$. This value may be compared to previous measurements of $b$. \citet{Jolly_etal_1998} measured $0.8\pm0.05$\,km\,s$^{-1}$ from CO absorption (they also measured a high $b$ value of 4.2\,km\,s$^{-1}$ from \ion{C}{1} absorption, potentially caused by modelling difficulties because the line is saturated). \citet{Roberge_etal_2000} obtained $1.3\pm0.5$\,km\,s$^{-1}$ from \ion{C}{1} absorption and $1.3\pm0.1$\,km\,s$^{-1}$ from CO absorption. \citet{Cataldi_etal_2014} and \citet{Nilsson_etal_2012} previously adopted 1.5\,km\,s$^{-1}$ for their models based on measurements of \ion{Ca}{2}\,K absorption \citep{Crawford_etal_1994}. The $b$ value derived from the ALMA data seems thus generally lower than in previous publications. %\citet{Vidal-Madjar_etal_2017} also find a smaller $b$ of $0.59^{+0.04}_{-0.01}$\,km\,s$^{-1}$ for \ion{Fe}{1} absorption in the ground base level; for absorption from excited levels on the other hand, they measure $b=1.01\pm0.06$\,km\,s$^{-1}$.

\section{Details of the deprojection procedure}\label{sec:deprojection_details}
The pv diagram can be used to get a deprojected view of the emission in the $(x,y)$ plane. As is discussed in appendix C of \citet{Matra_etal_2017}, a radius $r=(GM_*x^2/v^2)^{1/3}$ can be assigned to points $(x,v)$ of the pv diagram for an edge-on disk and circular Keplerian orbits. From this, we can find the position along the line of sight $y=\pm\sqrt{r^2-x^2}$. Note that for some points $(x,v)$ of the pv diagram, the term under the square root becomes negative. This simply means that for the given $x$, the radial velocity $v$ cannot be reached anywhere along the line of sight, i.e.\ we have $|v|>v_\mathrm{max}$, where $v_\mathrm{max}$ is the orbital velocity of the orbit with $r=x$. One has to choose how to distribute the flux from a given pv point among the two possible $y$ points (in front and behind the sky plane).

In practice, it is easiest to take an inverse approach. First, we decide to place all flux in front of the sky plane (i.e.\ we only consider $y<0$). Then, for a given point $(x,y)$, we calculate $r=\sqrt{x^2+y^2}$. From this, $v=-\sqrt{\frac{GM_*}{r}}\frac{x}{r}$, where the minus sign accounts for the known rotation sense of the $\beta$~Pic disk. We then look up the flux at $(x,v)$ in the pv diagram and assign it to the point $(x,y)$ in the deprojection.

In order to get a deprojection with a straightforward interpretation in terms of the distribution of the flux, it is also necessary to transform the flux units from W\,m$^{-2}$\,Hz$^{-1}$\,rad\,$^{-1}$ (as in the pv diagram, see figure \ref{fig:pv_diagram}) to W\,m$^{-2}$\,sr$^{-1}$. To do this, it is helpful to see the deprojection as a coordinate transformation from $(x,v)$ to $(x,y)$. The Jacobian determinant of this transformation (for $y=-\sqrt{r^2-x^2}$ as in figure \ref{fig:deprojection}) reads
\begin{equation}\label{eq:jacobian}
J = \frac{3}{2}\sqrt{GM_*}xy(x^2+y^2)^{-7/4}
\end{equation}
The flux read from the pv diagram is then multiplied by $|J|$ and an additional constant factor $\frac{\nu_0}{c}d$ (with $\nu_0$ the central frequency of the line and $d$ the distance between the observer and $\beta$~Pic).

\begin{figure}
\plotone{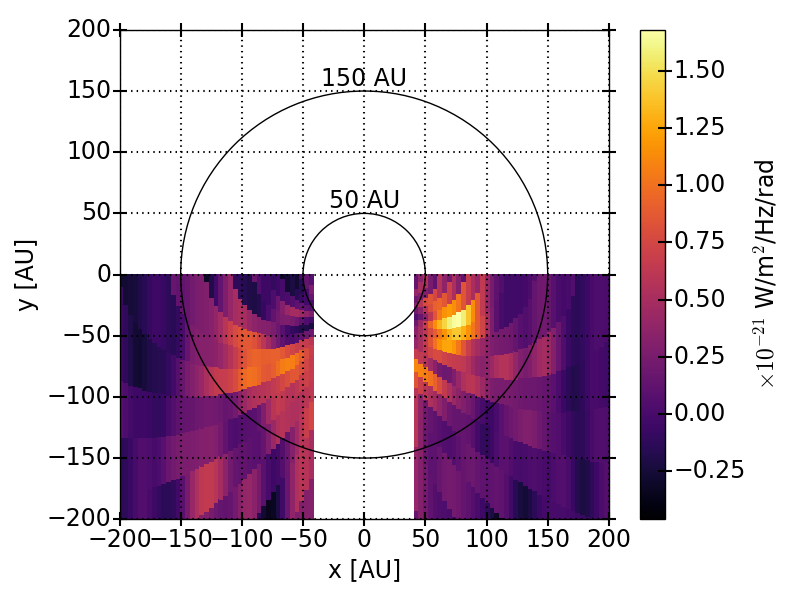}
\caption{Same as figure \ref{fig:deprojection}, but in the same units as the pv diagram, i.e. without multiplying the deprojection by the Jacobian shown in equation \ref{eq:jacobian}.\label{fig:deprojection_nojacobian}}
\end{figure}

If the flux units of the deprojection are not transformed \citep[as for example in][]{Matra_etal_2017}, an interpretation of the deprojection is not straightforward because the flux in a certain area of the deprojection does not equal the integral over $x$ and $y$ over this area. Figure \ref{fig:deprojection_nojacobian} shows the deprojection without multiplication by the Jacobian (i.e.\ in the same units as the pv diagram). Comparing to figure \ref{fig:deprojection}, it becomes apparent that a deprojection without transformed units might lead to visual mis-interpretation, for example about the relative amount of flux at large radii (say beyond 150\,au). Indeed, from equation \ref{eq:jacobian}, we see that the Jacobian gets smaller at large radii.

\section{Approximate calculation of the statistical equilibrium}\label{sec:approx_SE}
We solve the SE under the simplifying assumption that the photon escape fraction is high, i.e.\ we neglect (de)excitation by emitted line photons and assume that the background radiation field (CMB, stellar field and dust continuum) is not attenuated by the gas. In this section, we justify these assumptions.

The SE equation for an atomic level $i$ can be written
\begin{equation}
\sum_{j\neq i}\left(x_j(A_{ji}+C_{ji}+\bar{J}B_{ji}) - x_i(A_{ij}+C_{ij}+\bar{J}B_{ij})\right) = 0
\end{equation}
where $x_j$ is the fractional population of level $j$, $C_{ij}=K_{ij}n_\mathrm{e}$ is the collisional (de)excitation rate (with $K_{ij}$ the collisional rate coefficient and $n_\mathrm{e}$ the electron density), $\bar{J}$ the frequency-integrated mean intensity and $A_{ij}$ and $B_{ij}$ are the Einstein coefficients (where we set $A_{ij}=0$ if $i<j$). If we can show that $A_{ji}+C_{ji}\gg\bar{J}B_{ji}$ (for any $i,j$), then we have demonstrated that background radiation and line photons are not important to solve the SE.

The frequency-integrated mean intensity is the sum of the line emission and background radiation at each point of the disk: $\bar{J}=\bar{J}_\mathrm{line}+\bar{J}_\mathrm{backg}$. To get insight into the individual contribution of the two components, we consider them separately. First, we calculate $R_\mathrm{backg}=\bar{J}_\mathrm{backg}B_{ji}/(A_{ji}+C_{ji})$ for all transitions and all locations (except where the gas density is below 5\% of the peak density) for the best-fitting models described in sections \ref{sec:ring}, \ref{sec:ring_clump} and \ref{sec:eccentric_gas_distribution}. We find that $R_\mathrm{backg}<0.02$ for \ion{C}{1} and $R_\mathrm{backg}<0.001$ for \ion{C}{2}. Thus, although we included background radiation in our calculation, it is actually negligible. We double-check by re-computing the models without background radiation and find that the results indeed do not change.

Next, we consider (de)excitation by line emission and compute $R_\mathrm{line}$. To this end, we compute, for every location (again, except where the gas density is below 5\% of the peak density), the number of line photons arriving from the other grid points, neglecting optical depth and the velocity field (the velocity field could red/blue-shift photons from other grid points out of the transition). We are thus calculating an upper limit on $\bar{J}_\mathrm{line}$. We find that $R_\mathrm{line}<0.4$ for \ion{C}{1} and $R_\mathrm{line}<0.03$ for \ion{C}{2}. Thus, a more sophisticated model should include the full radiative transfer for \ion{C}{1}, but neglecting the line photons is a decent approximation given the quality of our data, since it considerably simplifies the calculation of the models.

As an additional test, we used the \texttt{LIME} code version 1.9.1 \citep{Brinch_Hogerheijde_2010} to calculate the full non-LTE radiative transfer (neglecting background radiation except for the CMB) for our best-fit models from sections \ref{sec:ring}, \ref{sec:ring_clump} and \ref{sec:eccentric_gas_distribution}. We find that the total flux computed by \texttt{LIME} differs at most by a factor 1.2 for both \ion{C}{1} and \ion{C}{2}.

\section{Gas density for eccentric orbits}\label{sec:derivation_eccentric_orbits}
In this section, we derive the gas surface density of the model with eccentric orbits presented in section \ref{sec:eccentric_gas_distribution}. We assume that all orbits have a common pericentre and that the distribution of eccentricities is known and given by $P(e)$. We search the probability $P(r,\theta)$ (which we assume is proportional to the surface density) to find a particle at radius $r$ and true anomaly $\theta$. We first consider $P(e,\theta)=P(e)P(\theta|e)$. In our model, a given eccentricity $e$ corresponds to a single orbit (because all orbits have a common pericentre). Thus, $P(\theta|e)$ is interpreted as the probability to find a particle at true anomaly $\theta$ \emph{along the orbit with eccentricity $e$}. The time a particle spends at a given $\theta$ is inversely proportional to the orbital velocity. Thus, we have
\begin{equation}
P(\theta|e)=\frac{C(e)}{|v(e,\theta)|}
\end{equation}
where the normalisation constant $C(e)=\left(\int\frac{1}{|v(e,\theta')|}\mathrm{d}\theta'\right)^{-1}$. The orbital velocity is given by
\begin{equation}
v(e,\theta) = \sqrt{\mu\frac{1+2e\cos\theta+e^2}{q(1+e)}}
\end{equation}
with $\mu=GM_*$ and $q$ the fixed pericentre distance. Next, we consider the transformation from $(e,\theta)$ to $(r,\theta)$ where
\begin{equation}\label{eq:r_e_theta}
r=\frac{q(1+e)}{1+e\cos\theta}
\end{equation}
The transformation is bijective, except for $\theta=0$. Therefore, we can write
\begin{equation}
P(r,\theta)=P(e,\theta)|J|
\end{equation}
with $J$ the Jacobian determinant of the transformation, given by $J=\frac{\partial e}{\partial r}$, which can be calculated from equation \ref{eq:r_e_theta}. At $\theta=0$, the density $P(r,\theta)$ diverges. In practice, this is easily handled by simply not sampling the singularity on the numerical grid.

\bibliographystyle{aasjournal}
\bibliography{bibliography}


%% Include this line if you are using the \added, \replaced, \x

\end{document}